\documentclass{emulateapj}

\usepackage{apjfonts}
\usepackage{natbib}
\usepackage{lscape}

\shorttitle{LARGE POPULATION OF AGN IN CLUSTERS} 
\shortauthors{MARTINI ET AL.}

\slugcomment{ApJ accepted [16 February 2006]}

\newcommand{\eg}{{\rm e.g.}}
\newcommand{\etal}{{\rm et al.}}
\newcommand{\ergs}{erg s$^{-1}$}
\newcommand{\kms}{km s$^{-1}$}
\newcommand{\chandra}{{\it Chandra}}
\newcommand{\oiii}{[\ion{O}{3}]}
\newcommand{\oii}{[\ion{O}{2}]}

\newcommand{\ha}{H$\alpha$}
\newcommand{\hb}{H$\beta$}
\newcommand{\hd}{H$\delta$}

\begin{document}

\title{Spectroscopic Confirmation of a Large AGN Population in Clusters of Galaxies} 

\author{Paul Martini\altaffilmark{1,2}, Daniel D. Kelson\altaffilmark{3}, 
Eunhyeuk Kim\altaffilmark{1}, 
John S. Mulchaey\altaffilmark{3}, and Alex A. Athey\altaffilmark{3}} 

\altaffiltext{1}{Harvard-Smithsonian Center for Astrophysics; 60 Garden Street; 
Cambridge, MA 02138} 
\altaffiltext{2}{Current Address: Department of Astronomy, The Ohio State 
University, 140 West 18th Avenue, Columbus, OH 43210, 
martini@astronomy.ohio-state.edu}
\altaffiltext{3}{Carnegie Observatories, 813 Santa Barbara St., Pasadena, CA 91101-1292} 

\begin{abstract}

We have completed a spectroscopic survey of X-ray point sources in eight 
low-redshift clusters of galaxies ($0.05 < z < 0.31$) and have identified 40 
cluster members with broad-band (0.3--8 keV) X-ray luminosities between 
$L_X = 8\times10^{40}$ and $4\times10^{43}$ \ergs. There are between two and 
ten X-ray sources per cluster. 
We use visible-wavelength emission lines, X-ray spectral shapes, and 
multiwavelength flux ratios to determine that at least 35 of these galaxies 
are Active Galactic Nuclei (AGN). 
From our spectroscopic survey of other candidate cluster members we estimate 
that the AGN fraction $f_A$ is $\sim 5$\% for cluster galaxies more luminous 
than $M_R = -20$ mag hosting AGN with broad-band X-ray luminosities above 
$L_X = 10^{41}$ \ergs, or $f_A(M_R<-20;L_X>10^{41}) \sim 5$\%. 
We stress that additional, lower-luminosity AGN are expected to be present 
in the $M_R < -20$ mag cluster members. 
Our data unambiguously demonstrate that cluster galaxies host AGN more 
frequently than previously expected. Only four of these galaxies have 
obvious visible-wavelength AGN signatures, even though their X-ray 
luminosities are too high for their X-ray emission to be due to populations 
of low-mass X-ray binaries or hot, gaseous halos. 
We attribute the significant difference in visible and X-ray AGN 
identification to dilution of low-luminosity AGN spectral signatures by host 
galaxy starlight and/or obscuration of accretion onto the central, supermassive 
black hole. 

\end{abstract}

\keywords{galaxies: active -- galaxies: clusters: general 
-- galaxies: evolution -- X-rays: galaxies -- X-rays: galaxies: clusters -- 
X-rays: general
}

\section{Introduction}

One of the surprising results of the many deep \chandra\ {\it X-ray 
Observatory} observations of clusters of galaxies has been growing evidence 
for higher surface densities of point sources toward clusters of galaxies 
relative to blank field observations. 
Early evidence for this excess point source density was found with 
measurements of cumulative source counts relative to the expectations 
from deep, blank field exposures \citep{cappi01,sun02,molnar02}. 
While the natural interpretation of these sources was that they are 
associated with the clusters, most of these sources were sufficiently 
bright that they would have to be Active Galactic Nuclei (AGN) if they 
were cluster members. 

This conclusion was unexpected because AGN, or any emission-line galaxies, are 
rarely identified in clusters. In a study of elliptical galaxies, 
\citet{osterbrock60} noted that \oii\ emission was only observed from 
isolated ellipticals or those in small groups and not from any of the 25 
members of rich clusters that were observed. \citet{gisler78} significantly 
improved the statistical significance of this result with a survey of over 
1300 galaxies. He found that the decrease in the number of emission-line 
galaxies in clusters was true of lenticular and spiral galaxies, as well as 
ellipticals, and presented the first tentative evidence that AGN may be 
found less commonly in rich clusters than the field. This result was 
confirmed with a subsequent, more uniformly-selected spectroscopic survey of 
1095 cluster galaxies and 173 field galaxies by \citet{dressler85}. These 
authors found that 7\% of cluster galaxies have emission-line nuclei, 
compared to 31\% of field galaxies. AGN were similarly found to be about a 
factor of five times more rare in clusters: only 1\% of cluster galaxies were 
AGN compared to 5\% of the field sample. 

The smaller AGN population in rich clusters relative to the field has been 
ascribed to differences in the frequency of AGN fueling episodes and the 
smaller numbers of cluster galaxies with significant reservoirs of cold gas 
\citep[e.g.][]{giovanelli85}. One of the most popular mechanisms for fueling 
AGN is the merger of two gas-rich galaxies as the strong gravitational torques 
from a major merger could provide sufficient gas inflow to the central, 
supermassive black hole to produce an AGN \citep[e.g.][]{barnes92}. However, 
the high velocity dispersions of rich clusters results in a low merger 
frequency because the galaxies' relative velocities are too high to form 
bound pairs. If galaxy mergers and significant reservoirs of cold gas are 
the most important parameters for fueling AGN, these arguments provide a 
natural explanation for the lower AGN fraction in clusters compared to the 
field. 

These differences in the galaxy population between rich and poor environments 
suggest that cluster AGN are a valuable probe of AGN and galaxy evolution. 
In \citet{martini02}, we measured redshifts for all of the 
X-ray sources with bright visible-wavelength counterparts in the field of the 
$z=0.15$ cluster Abell~2104. 
This work confirmed for the first time that many of the 
X-ray point sources were in fact cluster members and demonstrated that there 
may be a sufficiently large AGN population in clusters of galaxies for 
demographic studies. Particularly, we found six X-ray sources associated with 
cluster members, or approximately 5\% of all of the luminous cluster galaxies. 
Since the \citet{martini02} study, several other groups have carried out 
X-ray studies of clusters with either new spectroscopic observations or data 
obtained from the literature.
These observations include the identification of two galaxies in the $z=0.83$ 
cluster MS1054-0321 \citep{johnson03}, eight in the $z=0.08$ cluster 
Abell~2255 \citep{davis03}, six galaxies in the $z=0.06$ cluster Abell~3128 
\citep{smith03}, a recent study of the Coma cluster \citep{finoguenov04}, 
and one in the $z=0.59$ cluster MS2053-04 \citep{tran05}. 
The X-ray emission from many of these galaxies appears to be due to AGN based 
on their X-ray luminosity and/or X-ray spectral shape. The two sources in 
MS1054-0321 and the brightest cluster galaxy in MS2053-04 have $L_X > 10^{42}$ 
\ergs\, although only a small fraction of the lower-redshift X-ray sources 
have such high luminosities. For these cases it is possible that the 
X-ray emission is due to either low-mass X-ray binaries (LMXBs) or halos of 
hot gas.  Several recent visible-wavelength \citep{miller03a} and radio surveys 
\citep{miller03b,best04} have also identified AGN in clusters of galaxies and 
studied the environmental dependence of AGN. 

New studies of \chandra\ observations of larger samples of additional 
clusters have identified more examples of X-ray source overdensities 
\citep{cappelluti05,ruderman05}, although this is not true of every study 
\citep{kim04b}. 
The largest of these studies to date is the investigation by \citet{ruderman05} 
of the fields of 51 massive clusters from the MACS sample \citep{ebeling01}. 
These authors measured significant X-ray source overdensities within 3.5~Mpc 
of the cluster cores, and in particular a high concentration of sources within 
the central 0.5~Mpc. If the X-ray sources in the moderate to high-redshift 
clusters studied by \citet{cappelluti05} and \citet{ruderman05} are associated 
with cluster galaxies, their high inferred luminosities makes them very 
strong AGN candidates. 

At progressively lower X-ray luminosities, an AGN classification based on 
X-ray luminosity alone becomes less secure because populations of 
LMXBs, hot gaseous halos, and massive starbursts also produce significant 
X-ray emission. Redshift surveys of galaxies detected in deep X-ray 
observations have not always detected AGN spectral signatures 
\citep[\eg][]{mushotzky00,hornschemeier01,brandt05}. 
\citet{martini02} ascribed the absence 
of clear AGN signatures from even relatively low-redshift ($z \sim 0.15$) 
cluster X-ray sources to a combination of dilution by host galaxy starlight 
and obscuration by gas and dust. In the absence of clear AGN spectral 
signatures, multiwavelength spectral shape is the clearest indicator of the 
physical origin of the X-ray emission. 
The X-ray and visible-wavelength properties of resolved LMXB populations 
have now been carefully measured with \chandra\ for a large number of nearby, 
luminous elliptical galaxies \citep{kim04c} and have been observed to 
produce broad band (0.3--8 keV) X-ray luminosities as high as $10^{41}$ \ergs. 
Galaxies with X-ray emission from LMXBs follow a fairly tight relation 
between $B-$band and X-ray flux because the number of LMXBs approximately 
scales with the stellar mass of the host galaxy. 
Hot gaseous halos from large, early-type galaxies can also produce significant 
X-ray emission. While the intracluster medium may strip these halos, 
sources with soft X-ray luminosities of $10^{40} - 
10^{41}$ \ergs\ have been observed in many nearby clusters with \chandra\ 
\citep[e.g.][]{vikhlinin01,yamasaki02,sun05}. There is also a relation 
between X-ray and stellar luminosity for emission from hot gas, 
although with larger scatter \citep{canizares87} and the larger scatter 
is likely due to environmental influences \citep{brown00}. 
Finally, the most powerful, local starburst galaxies such as Arp~220 and 
NGC~3256 can have X-ray luminosities as high as $10^{42}$ \ergs\ due to 
hot gas associated with the starburst \citep{moran99,iwasawa99}, although 
these galaxies also have luminous, visible-wavelength emission lines and are 
expected to be rare in rich clusters. 
In addition to their multiwavelength properties, resolution is another 
discriminant between LMXBs, hot gas, star formation, and AGN. The first 
three mechanisms will all produce 
extended X-ray emission and should be resolved in \chandra\ observations 
of low-redshift clusters. 

Since our initial work on Abell~2104, we have completed a spectroscopic 
study of seven additional, low-redshift ($0.05 < z < 0.3$) clusters of 
galaxies. 
We have measured redshifts for the visible-wavelength counterparts of all of 
the X-ray sources in the fields of these clusters more luminous than 
approximately one magnitude fainter than $M_R^*$ and identified a total of 
40 X-ray sources in eight clusters. 
In this paper we describe their multiwavelength properties, conclude 
that most of these sources must be low-luminosity AGN, and demonstrate how 
the X-ray sources in these clusters can account for the excess point source 
surface densities. 
The observations and data analysis are described next in \S\ref{sec:obs}, 
while their multiwavelength properties are analyzed in section \S\ref{sec:res}. 
The results of this study, in particular the nature of the X-ray emission, 
calculation of the AGN fraction, and implications for the X-ray source 
overdensities observed toward clusters are discussed in \S\ref{sec:dis}.
We summarize this work in the last section. In future papers we will 
investigate the relationship of these X-ray sources to the cluster galaxy 
population and perform a detailed sensitivity analysis to derive the X-ray 
luminosity function in clusters of galaxies. 

\section{Observations and Measurements} \label{sec:obs}

\begin{deluxetable*}{lccccrcrrrrrc}
\tablecolumns{13}
\tablenum{1}
\tabletypesize{\scriptsize}
\tablecaption{Observation Log\label{tbl:obslog}}
\tablehead{
\colhead{Cluster} &
\colhead{Redshift} &
\colhead{RA} &
\colhead{DEC} &
\colhead{$N_H$} &
\colhead{ObsID} &
\colhead{Camera} &
\colhead{T [ks]} &
\colhead{Filters} &
\colhead{Imaging Date} &
\colhead{Spectra Date} &
\colhead{$N_X$} &
\colhead{$L_{X,lim}$} \\
\colhead{(1)} &
\colhead{(2)} &
\colhead{(3)} &
\colhead{(4)} &
\colhead{(5)} &
\colhead{(6)} &
\colhead{(7)} &
\colhead{(8)} &
\colhead{(9)} &
\colhead{(10)} &
\colhead{(11)} &
\colhead{(12)} &
\colhead{(13)} \\
}
\startdata
Abell 3125 & 0.059 & 03:27:22 & -53:30:00 & 1.6 & 892 & ACIS-I & 9 & $BVR$  & 2002 Sep & 2003 Oct &  6 & $5.3 \times 10^{40}$ \\
Abell 3128 & 0.060 & 03:30:24 & -52:32:00 & 1.5 & 893 & ACIS-I &19 & $BVR$  & 2002 Sep & 2003 Oct & 10 & $2.3 \times 10^{40}$ \\
Abell  644 & 0.070 & 08:17:25 & -07:30:42 & 6.4 &2211 & ACIS-I &29 & $BVR$  & 2003 Mar & 2003 Mar &  2 & $2.2 \times 10^{40}$ \\
Abell 2104 & 0.155 & 15:40:07 & -03:17:24 & 8.9 & 895 & ACIS-S &49 & $BVRI$ & 2003 Mar & 2002 Apr &  6 & $5.4 \times 10^{40}$ \\
Abell 1689 & 0.183 & 13:11:30 & -01:20:10 & 1.8 &1663 & ACIS-I &10 & $BVRI$ & 2003 Mar & 2003 Mar &  2 & $5.0 \times 10^{41}$ \\
Abell 2163 & 0.203 & 16:15:46 & -06:08:55 & 12.0&1653 & ACIS-I &70 & $BVR$  & 2003 Mar & 2003 Mar &  3 & $1.1 \times 10^{41}$ \\
MS1008     & 0.301 & 10:10:32 & -12:39:32 & 6.7 & 926 & ACIS-I &41 & $BVRI$ & 2003 Mar & 2003 Mar &  5 & $4.4 \times 10^{41}$ \\
AC 114     & 0.312 & 22:58:49 & -34:48:09 & 1.3 &1562 & ACIS-S &72 & $BVRI$ & 2002 Sep & 2002 Oct &  6 & $1.4 \times 10^{41}$ \\
\enddata
\tablecomments{
Cluster sample, properties, observations data, and the number of confirmed
cluster members with X-ray sources. Columns are: (1) Cluster name; (2)
Cluster redshift; (3 and 4) RA and DEC in for epoch J2000; (5) Galactic
Neutral Hydrogen column density in units of $10^{20}$ cm$^{-2}$; (6)
ObsID of the archival \chandra\ data; (7) ACIS Camera; (8) Useful integration
time of the \chandra\ data; (8) Visible-wavelength filters; (9) Date of
visible-wavelength imaging; (10) Date of spectroscopic observations; (11)
Number of X-ray sources in the cluster; (12) Estimate of the broad band
luminosity limit of the observations for a cluster member in \ergs\ (see
Section~\ref{sec:xdata}).
}
\end{deluxetable*}

This sample of eight clusters was chosen because the clusters have publicly 
available \chandra\ ACIS data with sufficiently long exposures to detect 
X-ray sources as faint as $L_X \sim 10^{41}$ \ergs\ at the cluster redshift. 
All of these clusters are also observable from Las Campanas Observatory in 
Chile. 
The cluster sample, redshift, and basic data about the X-ray, 
visible-wavelength imaging, and spectroscopic observations are presented 
in Table~\ref{tbl:obslog}. 
The field of view of the survey in each cluster is set by the field size of 
the camera employed to obtain the X-ray data and is either 
$8.3' \times 8.3'$ (ACIS-S) or $16.9' \times 16.9'$ (ACIS-I). 
We obtained visible-wavelength imaging with the 2.5m du~Pont telescope, 
identified likely counterparts to X-ray sources in these fields, and then 
obtained spectroscopy with the 6.5m Magellan Clay Telescope to determine which 
X-ray counterparts belong to the clusters. 
This search identified a sample of 40 cluster members, whose coordinates, 
redshifts, and visible-wavelength properties are provided in 
Table~\ref{tbl:catalog}. We describe the X-ray, imaging, and spectroscopic 
observations in the next subsections. 

\begin{deluxetable*}{lcccccccccll}
\tablecolumns{12}
\tablenum{2}
\tabletypesize{\tiny} 
\tablecaption{Visible-Wavelength Data\label{tbl:catalog}}
\tablehead{
\colhead{ID} &
\colhead{CXOU XID} &
\colhead{$z$} &
\colhead{$R$} &
\colhead{$B-R$} &
\colhead{$V-R$} &
\colhead{$R-I$} &
\colhead{[OII]} &
\colhead{\hd} &
\colhead{Notes} &
\colhead{Flags} &
\colhead{Lit ID} \\
\colhead{(1)} &
\colhead{(2)} &
\colhead{(3)} &
\colhead{(4)} &
\colhead{(5)} &
\colhead{(6)} &
\colhead{(7)} &
\colhead{(8)} &
\colhead{(9)} &
\colhead{(10)} &
\colhead{(11)} &
\colhead{(12)} \\
}
\startdata
A3125-1  &  J032723.5-532535 & 0.0642  &      15.93 (0.03) &   1.78 (0.06) &   0.65 (0.05) &      ...      &           ...      &    ...      &           ...      &     Lit-z     &        D80-052  \\
A3125-2  &  J032754.7-532217 & 0.0583  &      15.10 (0.03) &   1.81 (0.06) &   0.69 (0.05) &      ...      &           ...      &    ...      &           ...      &      ...      &        2MASXJ03275473-5322169  \\
A3125-3  &  J032752.0-532610 & 0.0609  &      15.26 (0.03) &   1.82 (0.06) &   0.70 (0.05) &      ...      &           ...      &    ...      &           ...      &      ...      &        2MASXJ03275206-5326099  \\
A3125-4  &  J032724.8-532518 & 0.0628  &      15.26 (0.03) &   1.84 (0.06) &   0.67 (0.05) &      ...      &           ...      &    ...      &           ...      &      ...      &        ...                     \\
A3125-5  &  J032705.0-532140 & 0.0625  &      16.08 (0.03) &   1.78 (0.06) &   0.71 (1.05) &      ...      &        -12.0 (0.5) &    ...      &     \ha,[NII],[SII]  &      ...      &        ESO155-IG031            \\
A3125-6  &  J032823.6-533436 & 0.0630  &          ...      &      ...      &      ...      &      ...      &           ...      &    ...      &           ...      & Lit-z,Lit-pos &        ESO155-G037,RGC02-241   \\
A3128-1  &  J033018.6-522856 & 0.0549  &      15.51 (0.03) &   1.64 (0.06) &   0.48 (0.05) &      ...      &           ...      &    ...      &           ...      &      ...      &        CSRG0262                \\
A3128-2  &  J032941.4-522936 & 0.0586  &      17.20 (0.03) &   1.20 (0.06) &   0.36 (0.05) &      ...      &         -2.8 (0.5) &   1.1 (0.2) &           ...      &      ...      &        S03-6,APMUKS(BJ)B032816.70-523949.5  \\
A3128-3  &  J032931.1-522716 & 0.0586  &          ...      &      ...      &      ...      &      ...      &           ...      &    ...      &           ...      & Lit-z,Edge    &        D80-135  \\
A3128-4  &  J033051.0-523031 & 0.0571  &      15.04 (0.03) &   1.51 (0.06) &   0.59 (0.05) &      ...      &           ...      &    ...      &           ...      &      ...      &        S03-3,2MASXJ03305107-5230315  \\
A3128-5  &  J033046.0-522335 & 0.0588  &          ...      &      ...      &      ...      &      ...      &           ...      &    ...      &           ...      & Lit-pos,$\Delta r$ &   2MASXJ03304589-5223385  \\
A3128-6  &  J033017.3-523407 & 0.0544  &      16.96 (0.03) &   1.28 (0.06) &   0.33 (0.05) &      ...      &           ...      &    ...      &           ...      &      ...      &        S03-5                   \\
A3128-7  &  J033013.6-523730 & 0.0648  &      15.15 (0.03) &   1.41 (0.06) &   0.24 (0.05) &      ...      &           ...      &    ...      &           ...      &      ...      &        S03-4,2MASXJ03301366-5237304  \\
A3128-8  &  J032950.6-523447 & 0.0643  &      15.31 (0.03) &   1.64 (0.06) &   0.48 (0.05) &      ...      &           ...      &    ...      &           ...      & $\Delta r$    &        2MASXJ03295060-5234471  \\
A3128-9  &  J033039.3-523206 & 0.0622  &      16.50 (0.03) &   1.63 (0.06) &   0.46 (0.05) &      ...      &           ...      &    ...      &           ...      &      ...      &        2MASXJ03303924-5232066  \\
A3128-10 &  J033038.4-523710 & 0.0599  &      14.94 (0.03) &   1.54 (0.06) &   0.59 (0.05) &      ...      &           ...      &    ...      &           ...      & $\Delta r$    &        S03-2,2MASXJ03303848-5237096  \\
A644-1   &  J081739.6-073309 & 0.0726  &      16.90 (0.03) &   0.95 (0.06) &   0.35 (0.05) &      ...      &           ...      &    ...      &     NLS1           &      ...      &        ...                     \\
A644-2   &  J081748.1-073732 & 0.0783  &      15.93 (0.03) &   1.81 (0.06) &   0.66 (0.05) &      ...      &           ...      &   2.0 (0.3) &           ...      &      ...      &        ...                     \\
A2104-1  &  J154023.6-031347 & 0.159   &      17.30 (0.03) &   2.43 (0.06) &   0.81 (0.05) &   0.98 (0.05) &           ...      &    ...      &     S2             &      ...      &        MKMT02-1                \\
A2104-2  &  J154016.7-031507 & 0.161   &      18.65 (0.03) &   2.32 (0.06) &   0.78 (0.05) &   0.85 (0.05) &           ...      &    ...      &           ...      &      ...      &        MKMT02-2                \\
A2104-3  &  J154009.4-031519 & 0.155   &      17.16 (0.03) &   2.46 (0.06) &   0.79 (0.05) &   0.85 (0.05) &           ...      &    ...      &           ...      &      ...      &        MKMT02-3                \\
A2104-4  &  J154014.0-031704 & 0.157   &      19.60 (0.03) &   2.22 (0.06) &   0.69 (0.05) &   0.78 (0.05) &           ...      &    ...      &           ...      & $\Delta r$    &        MKMT02-4                \\
A2104-5  &  J154019.5-031825 & 0.162   &      19.56 (0.03) &   2.04 (0.06) &   0.67 (0.05) &   0.76 (0.05) &           ...      &    ...      &           ...      &      ...      &        MKMT02-5                \\
A2104-6  &  J154003.9-032039 & 0.154   &      17.77 (0.03) &   2.43 (0.06) &   0.74 (0.05) &   0.87 (0.05) &           ...      &    ...      &           ...      & $\Delta r$    &        MKMT02-6                \\
A1689-1  &  J131145.4-012336 & 0.187   &      17.75 (0.03) &   2.19 (0.06) &   0.77 (0.05) &   0.70 (0.05) &           ...      &    ...      &           ...      &      ...      &        APMUKS(BJ)B130911.16-010741.8  \\
A1689-2  &  J131135.6-012012 & 0.200   &      18.34 (0.03) &   1.41 (0.06) &   0.48 (0.05) &   0.50 (0.05) &           ...      &    ...      &     S2             & $\Delta z$    &        APMUKS(BJ)B130901.41-010417.1  \\
A2163-1  &  J161524.4-060904 & 0.201   &      17.76 (0.03) &   1.89 (0.06) &   0.70 (0.05) &      ...      &        -14.0 (1.0) &   2.5 (0.5) &     S2             &    ...   &        ...                     \\
A2163-2  &  J161548.9-061512 & 0.201   &      20.04 (0.03) &   2.24 (0.07) &   0.76 (0.05) &      ...      &           ...      &    ...      &           ...      & $\Delta r$    &        ...                     \\
A2163-3  &  J161543.6-061730 & 0.200   &      18.34 (0.03) &   2.52 (0.06) &   0.86 (0.05) &      ...      &           ...      &    ...      &           ...      &      ...      &        ...                     \\
MS1008-1 &  J101018.7-123744 & 0.298   &      19.60 (0.03) &   2.12 (0.06) &   0.71 (0.05) &   0.83 (0.05) &         -6.0 (0.5) &   2.9 (0.3) &     [OIII]          &      ...      &        PPP-001519              \\
MS1008-2 &  J101005.2-123834 & 0.305   &      21.53 (0.04) &   2.29 (0.10) &   0.77 (0.15) &   0.72 (0.08) &           ...      &    ...      &           ...      & $\Delta r$    &        ...                     \\
MS1008-3 &  J101035.3-124021 & 0.309   &      19.17 (0.03) &   2.65 (0.06) &   1.02 (0.05) &   0.96 (0.05) &           ...      &    ...      &           ...      &      ...      &        PPP-000760              \\
MS1008-4 &  J101026.5-123810 & 0.297   &      19.71 (0.03) &   2.27 (0.06) &   0.86 (0.05) &   0.84 (0.05) &           ...      &   0.7 (0.2) &           ...      &      ...      &        PPP-001427              \\
MS1008-5 &  J101032.3-123934 & 0.301   &      18.96 (0.03) &   2.46 (0.06) &   0.92 (0.05) &   0.88 (0.05) &           ...      &   0.8 (0.2) &           ...      &      ...      &        ...                     \\
AC114-1  &  J225852.9-344846 & 0.304   &      20.01 (0.03) &   1.83 (0.06) &   0.77 (0.05) &   0.47 (0.05) &           ...      &   8.0 (0.6) &           ...      &      ...      &        CN84-191                \\
AC114-2  &  J225851.4-344912 & 0.321   &      21.42 (0.04) &   2.09 (0.20) &   1.24 (0.20) &   0.74 (0.08) &           ...      &    ...      &           ...      &      ...      &        ...                     \\
AC114-3  &  J225849.3-344701 & 0.313   &      19.26 (0.03) &   2.01 (0.06) &   0.96 (0.05) &   0.69 (0.08) &         -3.0 (0.8) &    ...      &     \hb,[OIII],\ha  &      ...      &        CN84-087                \\
AC114-4  &  J225846.2-344945 & 0.318   &      20.98 (0.04) &   2.32 (0.15) &   1.16 (0.15) &   0.70 (0.08) &           ...      &   3.2 (0.1) &           ...      &      ...      &        CBB2001-0552            \\
AC114-5  &  J225857.4-345059 & 0.322   &      21.02 (0.05) &   2.30 (0.15) &   1.80 (0.15) &   1.50 (0.08) &           ...      &    ...      &           ...      &     Lit-z     &        CBB2001-0365            \\
AC114-6  &  J225842.0-344747 & 0.310   &      18.14 (0.03) &   2.68 (0.06) &   1.27 (0.05) &   0.79 (0.05) &           ...      &    ...      &           ...      &     Lit-z     &        CN84-003                \\
\enddata
\tablecomments{
Visible-Wavelength Data. Columns are: (1) ID used in this paper; (2) object
ID; (3) redshift (4) observed $R-$band magnitude and associated
uncertainty; (5--7) observed colors; (8) [OII] equivalent width (\AA);
(9) H$\delta$ equivalent width (\AA); (10) Notes on prominent emission lines;
(11) Identification flags; (12) Other object IDs from the literature.
The identification flags in column 11 are Lit-z (redshift obtained from
literature source only), Lit-pos (object position obtained from the
Digital Sky Survey), Edge (object is near the edge of our $R-$band image,
$\Delta z$ (object has a large redshift offset from the cluster mean),
and $\Delta r$ (object center is more than 1$\sigma$ outside the X-ray
error circle). References for the literature positions and/or redshifts
used for this study are A3125-1: \citet{caldwell97}; A3125-6:
\citet{dressler80,rose02}; A3128-3: \citet{caldwell97}; A3128-5:
\citet{dressler80}; AC114-5: \citet{couch01}; AC114-6: \citet{couch84}.
}
\end{deluxetable*}

  \subsection{X-ray Data} \label{sec:xdata} 

The X-ray observations were initially processed with a standard application 
of the CIAO pipeline and the {\it wavdetect} source identification tool, 
although they have since been reprocessed with a custom X-ray photometry 
package named XPROCES (E.~Kim \etal\ 2006, {\it in prep}). 
As this is the first application of the package we provide a brief description 
here, although defer a complete description to a future paper. 

The first step is to produce a new level 2 event file following the standard 
science threads, including suitable filtering to remove flare events and 
CTI correction. XPROCES then runs the {\it wavdetect} task on the unbinned 
images within the CIAO package (Version 3.2) to identify point 
sources. We set the detection threshold at $10^{-6}$, searched scale 
sizes of $<1~2~4~8~16~32~64>$ pixels to identify sources, and employed an 
exposure map for an energy of 1.5 keV. 
The {\it wavdetect} detection threshold of $10^{-6}$ corresponds to one false 
detection per $10^6$ pixels, or one per unbinned ACIS chip. We therefore expect 
between one (ACIS-S) and four (ACIS-I) false detections per cluster field. 
In the same area (1000x1000 \chandra\ pixels or $8.3' \times 8.3'$) in which 
we expect one false detection, there $\sim 200 R<21$ mag galaxies. 
For a generous matching tolerance of 
$2''$, the probability of a chance superposition between a false detection 
and a bright galaxy is less than 3\% for the entire survey area. This is an 
upper limit on the false detection probability, as we also have a strict 
signal-to-noise cut on the X-ray data that will eliminate many false 
detections. The probability that a false detection will be cross-identified 
with a cluster member is lower still because many bright galaxies are not 
cluster members. 

XPROCES prepares the event file for photometry with several additional 
steps. The first step is detection and removal of new hot pixels and cosmic 
ray afterglow events, followed by the standard charge transfer inefficiency and 
time-dependent gain correction. At this stage flare events are 
detected and removed as well. 
The properties of the sources are then determined with aperture photometry, 
where the aperture size is set by the 95\% encircled energy fraction of 
the PSF at 1.5 keV (the energy of maximum quantum efficiency) at a given 
off-axis angle. 
We measure source counts in five X-ray bands following 
the definitions used by \citet{kim04a}. These are the Broad (0.3-8 keV), 
Soft (0.3-2.5 keV), Hard (2.5-8 keV), $S_1$ (0.3-0.9 keV), and $S_2$ 
(0.9-2.5 keV) 
bands. The same aperture size is used for all bands even though there are 
slight changes in the size of the PSF with energy. This is a small source of 
error. For example, the 95\% encircled energy at 1.5 keV corresponds 
to the 92\% encircled energy at 3.5 keV. We do not use the 95\% encircled 
energy aperture at 3.5 keV because it would increase the potential for 
source overlap. 
The counts for each source in these five bands are presented in 
Table~\ref{tbl:xdata}. Only sources detected at greater than 
3$\sigma$ significance in the Broad band are included in this sample and all of 
these sources have at least seven counts in the Broad band. Count values are 
not shown if the source counts corresponded to less than a 3$\sigma$ detection 
in any given band. The background flux for each source has been measured within 
an annulus from two to five times the source radius and this value has been 
subtracted from the aperture measurement. We have calculated the uncertainties 
and the 3$\sigma$ significance with the \citet{gehrels86} approximation for 
small $N$. 

For each source we calculate several quantities to parametrize the X-ray 
spectral energy distribution. 
Following \citet{kim04a} we define the hardness ratio $HR = (H-S)/(H+S)$ and 
the colors $C21 = {\rm log} S_1/S_2$ and $C32 = {\rm log} S_2/H$. These 
quantities are commonly-used characterizations of spectral shape. 
X-ray quantiles are a recently developed measure of spectral shape that 
are based on the distribution of energy values in a spectrum \citep{hong04}. 
A quantile is defined such that if $N$\% of the counts in a spectrum are 
below some energy $E_{N\%}$, then the quantile is: 
\begin{equation}
Q_N \equiv \frac{E_{N\%} - E_{lo}}{E_{up} - E_{lo}}
\end{equation}
where $E_{lo}$ and $E_{up}$ are the lower and upper energy boundaries of the 
spectrum and are taken here to be 0.3 and 8.0 keV, respectively. 
The virtue of X-ray quantiles is that they are less susceptible to biases due 
the relative energy sensitivity of an instrument. 
Following \citet{hong04}, we have calculated the quantities 
$Q_X = 3 Q_{25}/Q_{75}$ and $Q_Y = {\rm log_{10}} Q_{50}/(1-Q_{50})$. 
The hardness ratio, X-ray colors, and quantiles for our sample are listed in 
Table~\ref{tbl:xdata}.

We calculate fluxes in the broad, soft, and hard bands with an energy
conversion factor calculated from the foreground Galactic extinction toward
each cluster \citep{stark92} and assume a $\Gamma = 1.7$ power-law typical
of AGN, where $\Gamma$ is the slope of the power-law photon flux density 
$N_E \propto E^{-\Gamma}$. The energy conversion factor is calculated for 
each chip at the maximum value of the exposure map and then scaled to the 
location of each source with the exposure map. This approach is similar to 
that described in \citet{kim04a}. 
We calculate the X-ray luminosity of each source based on the luminosity 
distance \citep[\eg][]{carroll92} to each cluster and assume a cosmology with 
$(\Omega_M, \Omega_\Lambda, h) = (0.3, 0.7, 0.7)$.
The fluxes are measured in the observed bandpasses and the luminosities have
been calculated for the restframe bandpass assuming a $\Gamma = 1.7$ model.
The source fluxes and luminosities are presented in Table~\ref{tbl:lum}.
For a $\Gamma = 1.7$ power-law plus Galactic absorption model we are complete
to broad-band luminosity limits of $L_X = 2 \times 10^{40}$ \ergs\ (A644
at $z = 0.07$) to $L_X = 5 \times 10^{41}$ \ergs (A1689 at $z=0.18$).
These luminosity limits are listed as the last column of Table~\ref{tbl:obslog}
and correspond to a five count detection on axis. 
Thirty eight of the 40 sources are unresolved, based on a 
comparison to the \chandra\ PSF at similar off-axis angle. The remaining two 
sources (AC114-1, AC114-2) are sufficiently embedded in the diffuse, 
intracluster gas that it is difficult to determine if the emission due to the 
galaxy is unresolved. 

\begin{deluxetable*}{lccccccccccc}
\tablecolumns{11}
\tablenum{3}
\tabletypesize{\tiny}
\tablecaption{X-Ray Photometry, Colors, and Quantiles\label{tbl:xdata}}
\tablehead{
\colhead{ID} &
\colhead{$B$} &
\colhead{$S$} &
\colhead{$H$} &
\colhead{$S_1$} &
\colhead{$S_2$} &
\colhead{$C21$} &
\colhead{$C32$} &
\colhead{$HR$} &
\colhead{$Q_X$} &
\colhead{$Q_Y$} \\
\colhead{(1)} &
\colhead{(2)} &
\colhead{(3)} &
\colhead{(4)} &
\colhead{(5)} &
\colhead{(6)} &
\colhead{(7)} &
\colhead{(8)} &
\colhead{(9)} &
\colhead{(10)} &
\colhead{(11)} \\
}
\startdata
A3125-1  &   21.2 (3.7) &   20.4 (3.6) &      ...     &    5.0 (1.5) &   15.4 ( 3.0) &   -0.5 (0.3) &       ...    &       ...    &   -0.9 (0.1) &    1.7 (0.8)\\
A3125-2  &   15.4 (2.8) &   16.0 (3.0) &      ...     &   11.2 (2.4) &    4.9 ( 1.3) &    0.4 (0.4) &       ...    &       ...    &   -1.1 (0.1) &    2.2 (0.4)\\
A3125-3  &    7.6 (1.8) &    8.3 (2.0) &      ...     &    4.8 (1.4) &    3.5 ( 1.1) &    0.1 (0.5) &       ...    &       ...    &   -1.1 (0.3) &    1.7 (0.6)\\
A3125-4  &    8.0 (1.9) &    6.3 (1.7) &      ...     &      ...     &    3.4 ( 1.1) &      ...     &       ...    &       ...    &   -0.9 (0.6) &    0.9 (0.6)\\
A3125-5  &   25.9 (4.1) &   11.9 (2.5) &   14.0 (2.8) &      ...     &   11.4 ( 2.5) &      ...     &   -0.1 (0.2) &    0.1 (0.3) &   -0.4 (0.1) &    1.5 (0.3)\\
A3125-6  &   13.7 (2.6) &   13.0 (2.7) &      ...     &    3.9 (1.2) &    9.1 ( 2.1) &   -0.4 (0.4) &       ...    &       ...    &   -1.0 (0.3) &    1.1 (0.6)\\
A3128-1  &   19.4 (3.4) &   17.9 (3.3) &      ...     &    9.6 (2.2) &    8.4 ( 2.0) &    0.1 (0.3) &       ...    &       ...    &   -1.1 (0.2) &    1.5 (0.8)\\
A3128-2  & 176.7 (12.2) & 124.2 (10.1) &   52.5 (6.2) &   39.0 (5.3) &   85.1 ( 8.2) &   -0.3 (0.1) &    0.2 (0.1) &   -0.4 (0.1) &   -0.7 (0.1) &    0.9 (0.1)\\
A3128-3  &   15.2 (2.5) &   11.2 (2.2) &    4.0 (1.0) &      ...     &    9.2 ( 2.0) &       ...    &    0.4 (0.5) &   -0.5 (0.4) &   -0.6 (0.3) &    1.3 (1.2)\\
A3128-4  &   98.2 (8.8) &   96.5 (8.8) &      ...     &   43.0 (5.6) &   53.5 ( 6.3) &   -0.1 (0.1) &       ...    &       ...    &   -1.0 (0.0) &    1.7 (0.1)\\
A3128-5  &   14.4 (2.3) &   10.2 (1.9) &    4.2 (1.0) &    4.8 (1.3) &    5.4 ( 1.2) &    0.0 (0.5) &    0.1 (0.6) &   -0.4 (0.5) &   -0.7 (0.4) &    0.6 (0.5)\\
A3128-6  & 216.8 (13.7) & 175.1 (12.2) &   41.8 (5.5) &   68.7 (7.3) &  106.4 ( 9.3) &   -0.2 (0.1) &    0.4 (0.1) &   -0.6 (0.1) &   -0.8 (0.1) &    0.9 (0.1)\\
A3128-7  &   33.9 (4.6) &   33.7 (4.7) &      ...     &   20.2 (3.6) &   13.5 ( 2.6) &    0.2 (0.2) &       ...    &       ...    &   -1.1 (0.1) &    1.7 (0.2)\\
A3128-8  &   26.8 (3.2) &   20.4 (2.7) &    6.4 (1.4) &    5.2 (1.2) &   15.2 ( 2.3) &   -0.5 (0.4) &    0.4 (0.4) &   -0.5 (0.3) &   -0.8 (0.2) &    1.1 (0.6)\\
A3128-9  &   25.4 (4.0) &   19.2 (3.4) &    6.2 (1.6) &    3.4 (1.1) &   15.9 ( 3.0) &   -0.7 (0.4) &    0.4 (0.3) &   -0.5 (0.2) &   -0.6 (0.2) &    1.2 (0.4)\\
A3128-10 &   38.7 (5.1) &   38.7 (5.2) &      ...     &   13.3 (2.8) &   25.4 ( 4.0) &   -0.3 (0.2) &       ...    &       ...    &   -1.0 (0.1) &    1.8 (0.2)\\
A644-1   & 250.4 (14.6) & 167.0 (11.8) &   83.4 (8.1) &   98.3 (8.9) &   68.7 ( 7.2) &    0.2 (0.1) &   -0.1 (0.1) &   -0.3 (0.1) &   -0.7 (0.1) &    0.4 (0.1)\\
A644-2   &   28.8 (3.8) &    6.1 (1.3) &   22.6 (3.5) &      ...     &    3.4 ( 0.8) &       ...    &   -0.8 (0.6) &    0.6 (0.3) &   -0.1 (0.2) &    1.6 (0.3)\\
A2104-1  &3778.3 (60.3) &1797.0 (41.3) &1981.3 (43.4) &   66.8 (7.0) & 1730.3 (40.5) &   -1.4 (0.1) &   -0.1 (0.0) &    0.0 (0.0) &   -0.4 (0.0) &    1.2 (0.0)\\
A2104-2  &   80.9 (7.8) &   22.5 (3.6) &   58.3 (6.6) &    3.7 (1.1) &   18.9 ( 3.3) &   -0.7 (0.4) &   -0.5 (0.1) &    0.4 (0.1) &   -0.2 (0.1) &    1.3 (0.1)\\
A2104-3  &   29.3 (4.1) &   29.1 (4.2) &      ...     &   14.3 (2.7) &   14.8 ( 2.8) &    0.0 (0.2) &       ...    &       ...    &   -1.1 (0.1) &    1.7 (0.2)\\
A2104-4  &   25.5 (3.3) &   26.9 (3.6) &      ...     &    6.3 (1.5) &   20.6 ( 3.1) &   -0.5 (0.3) &       ...    &       ...    &   -0.9 (0.2) &    1.6 (0.3)\\
A2104-5  & 166.1 (11.7) & 136.5 (10.5) &   29.5 (4.4) &   36.0 (5.0) &  100.5 ( 8.9) &   -0.4 (0.1) &    0.5 (0.1) &   -0.6 (0.1) &   -0.9 (0.1) &    1.2 (0.1)\\
A2104-6  &   28.6 (3.9) &   12.2 (2.2) &   16.4 (3.0) &      ...     &   13.0 ( 2.4) &       ...    &   -0.1 (0.2) &    0.1 (0.3) &   -0.3 (0.1) &    1.0 (0.4)\\
A1689-1  &   10.6 (2.2) &   10.4 (2.3) &      ...     &      ...     &    7.8 ( 1.9) &       ...    &       ...    &       ...    &   -0.8 (0.2) &    1.6 (0.6)\\
A1689-2  &   16.1 (2.8) &   11.6 (2.3) &    4.5 (1.2) &      ...     &   11.3 ( 2.3) &       ...    &    0.4 (0.4) &   -0.4 (0.4) &   -0.7 (0.5) &    0.9 (0.3)\\
A2163-1  & 159.4 (11.0) &   11.3 (1.9) & 148.1 (10.9) &    6.3 (1.5) &    5.0 ( 1.1) &    0.1 (0.5) &   -1.5 (0.4) &    0.9 (0.1) &    0.0 (0.0) &    1.9 (0.1)\\
A2163-2  &   35.8 (4.9) &   25.6 (4.1) &   10.1 (2.2) &      ...     &   23.0 ( 3.8) &       ...    &    0.4 (0.2) &   -0.4 (0.2) &   -0.6 (0.1) &    1.1 (0.2)\\
A2163-3  &   12.1 (2.3) &   12.5 (2.5) &      ...     &    3.3 (1.0) &    9.2 ( 2.1) &   -0.4 (0.5) &       ...    &       ...    &   -0.9 (0.1) &    1.9 (0.8)\\
MS1008-1 &    7.9 (1.9) &    4.5 (1.3) &    3.5 (1.1) &      ...     &      ...      &       ...    &       ...    &       ...    &   -0.6 (0.5) &    0.4 (0.7)\\
MS1008-2 &    8.1 (2.0) &    5.6 (1.5) &      ...     &      ...     &    5.8 ( 1.6) &       ...    &       ...    &       ...    &   -0.5 (0.7) &    0.8 (0.5)\\
MS1008-3 &   67.0 (6.9) &   52.1 (6.0) &   14.9 (2.9) &   12.9 (2.7) &   39.2 ( 5.1) &   -0.5 (0.2) &    0.4 (0.2) &   -0.6 (0.1) &   -0.7 (0.1) &    1.1 (0.3)\\
MS1008-4 &    8.9 (1.9) &    7.1 (1.6) &      ...     &      ...     &    5.7 ( 1.4) &       ...    &       ...    &       ...    &   -0.9 (0.5) &    1.1 (1.1)\\
MS1008-5 &   37.3 (3.3) &   29.2 (2.9) &    8.1 (1.4) &   10.3 (1.9) &   18.9 ( 2.1) &   -0.3 (0.3) &    0.4 (0.4) &   -0.6 (0.3) &   -0.8 (0.1) &    1.3 (0.5)\\
AC114-1  & 933.1 (29.0) & 910.9 (28.7) &   22.2 (3.4) & 706.1 (25.4) &  204.7 (12.8) &    0.5 (0.0) &    1.0 (0.1) &   -1.0 (0.0) &   -1.4 (0.0) &    0.8 (0.1)\\
AC114-2  &   47.6 (4.3) &   32.1 (3.3) &   15.5 (2.6) &    7.1 (1.1) &   25.0 ( 3.0) &   -0.6 (0.4) &    0.2 (0.2) &   -0.3 (0.2) &   -0.7 (0.2) &    0.7 (0.2)\\
AC114-3  &  122.0 (9.6) &  108.2 (9.0) &   13.8 (2.6) &   46.9 (5.7) &   61.3 ( 6.5) &   -0.1 (0.1) &    0.6 (0.2) &   -0.8 (0.1) &   -0.9 (0.1) &    1.1 (0.2)\\      AC114-4  &   63.2 (6.5) &   23.0 (3.4) &   40.2 (5.3) &      ...     &   22.9 ( 3.6) &       ...    &   -0.2 (0.1) &    0.3 (0.2) &   -0.1 (0.1) &    1.1 (0.1)\\          
AC114-5  &    7.8 (1.5) &    5.7 (1.2) &      ...     &    5.5 (1.3) &      ...      &       ...    &       ...    &       ...    &   -1.3 (1.5) &    0.6 (0.7)\\
AC114-6  &    7.5 (1.3) &    7.2 (1.3) &      ...     &    3.6 (0.9) &    3.6 ( 0.8) &    0.0 (0.7) &       ...    &       ...    &   -1.1 (1.3) &    0.7 (0.7)\\
\enddata
\tablecomments{
X-ray data and measurements. Columns (2) -- (6) contain the measured counts
in the (2) Broad (0.3-8 keV), (3) Soft (0.3-2.5 keV), (4) Hard (2.5-8 keV),
(5) $S_1$ (0.3-0.9 keV), and (6) $S_2$ (0.9-2.5 keV) energy bands. The  
X-ray colors C21 and C32 are provided in columns (7) and (8), the hardness
ratio HR in column (9), and the quantiles $Q_X$ and $Q_Y$ in columns
(10) and (11). See Section~\ref{sec:xdata} for further details.
}
\end{deluxetable*}

  \subsection{Ground-based Images} 

\begin{figure*}
\figurenum{1} 
\plotone{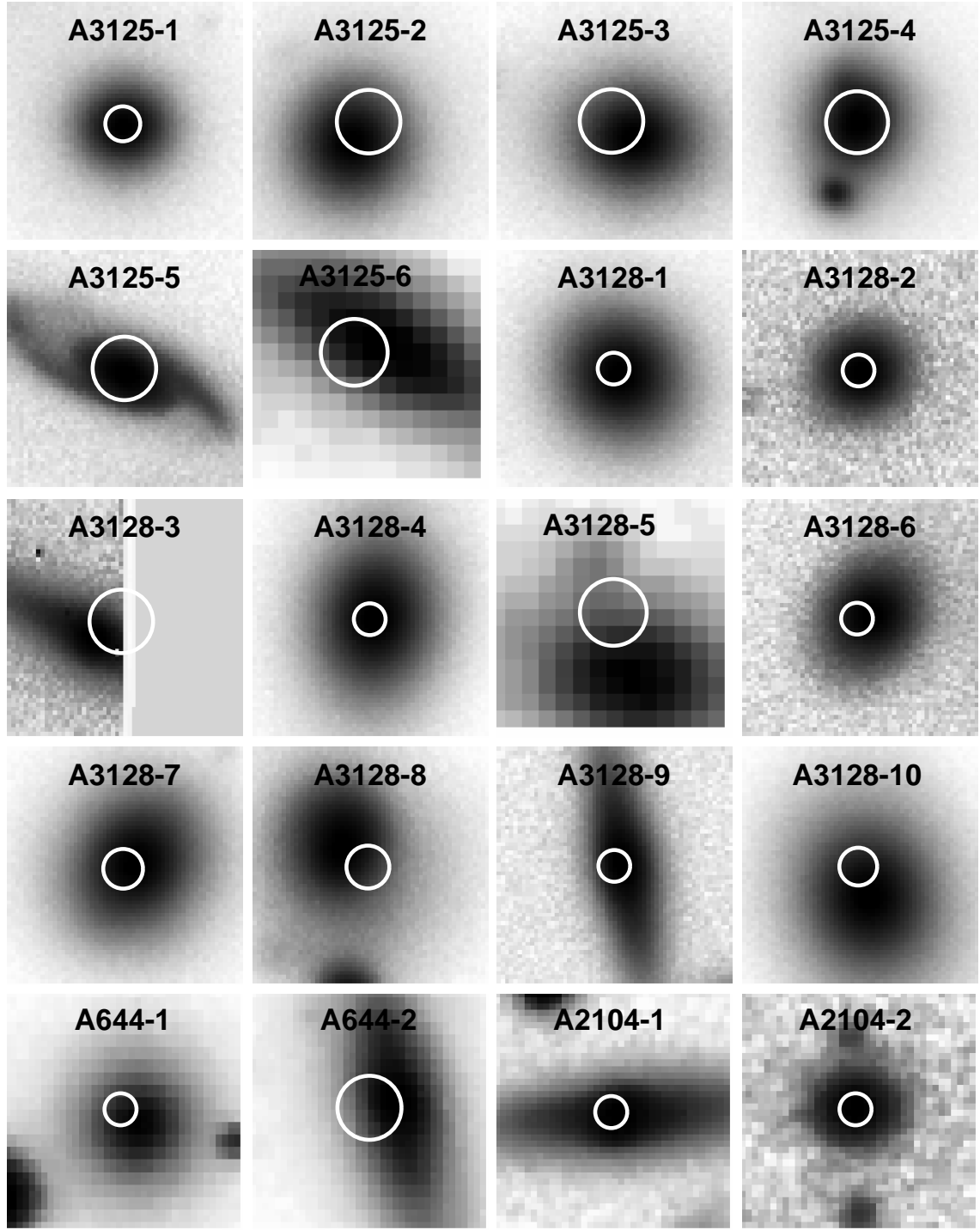} 
\caption{
\label{fig:charts} 
Cutouts from our R-band images of each cluster member along with 95\% 
X-ray error circles. 
A3125-6 and A3128-5 were obtained from the {\it Digitized Sky Survey}. 
Each panel is $15'' \times 15''$ and North is up and East is to 
the left. The panels are ordered as in Table~\ref{tbl:catalog}. 
}
\end{figure*}

\begin{figure*}
\figurenum{1} 
\plotone{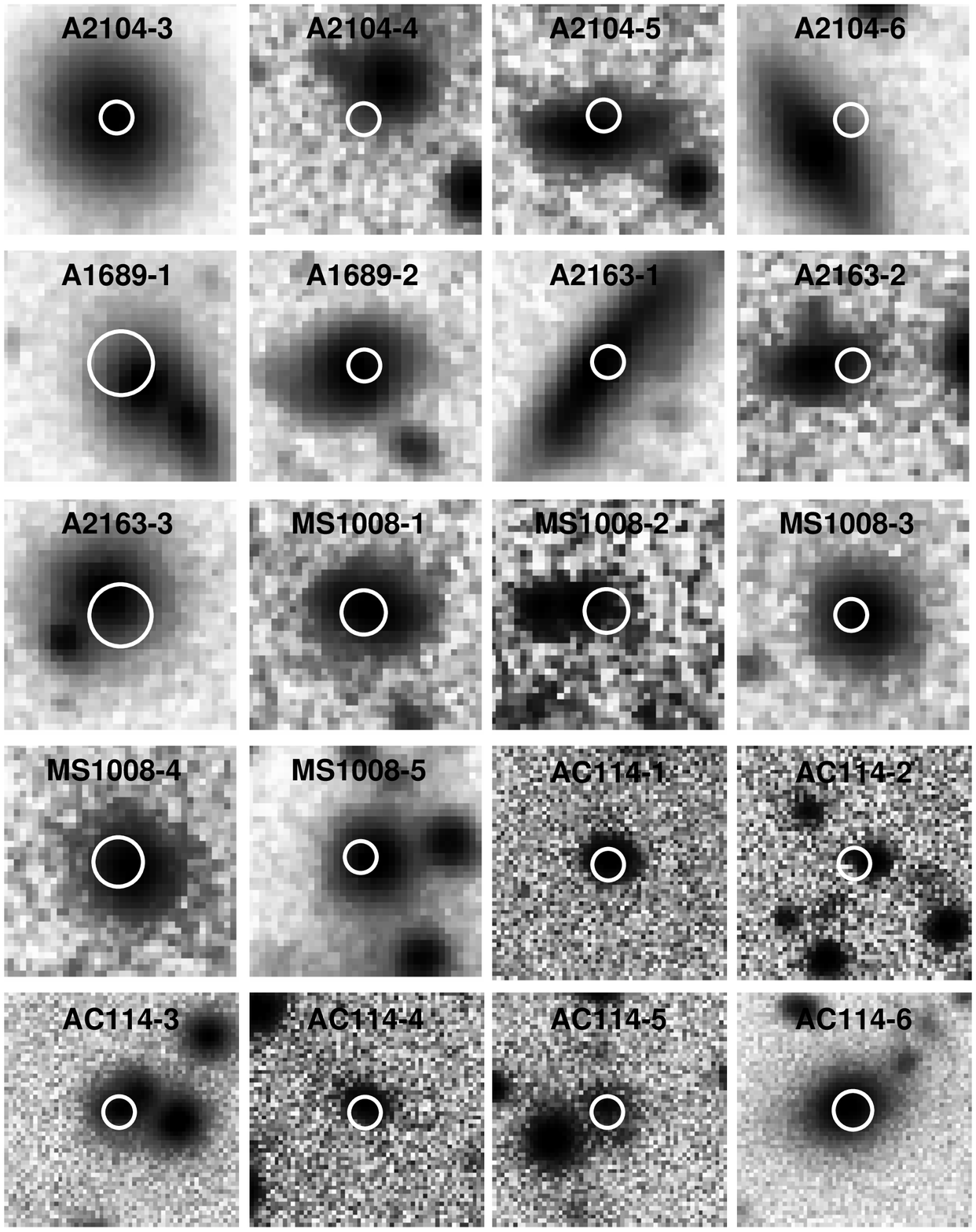} 
\caption{{\it Continued}}
\end{figure*}

We obtained ground-based $BVR$ photometry for the three $z < 0.1$ clusters 
and $BVRI$ photometry for the five higher-redshift clusters with the 2.5m 
du~Pont telescope at Las Campanas Observatory. 
The observations obtained in 2003 March employed the WFCCD camera, which has a 
plate scale of $0.775'' {\rm pix}^{-1}$, and the observations in 2002 September 
employed the TEK5 CCD camera, which has a plate scale of $0.385'' 
{\rm pix}^{-1}$. These observations encompass nearly all of the fields-of-view 
of the primary \chandra\ camera (ACIS-I or ACIS-S). For two clusters 
(A3125 and A3128) we obtained multiple pointings with the TEK5 camera to tile 
the \chandra\ field of view. These images were combined into single mosaics 
with the SWARP\footnote{http://terapix.iap.fr/rubrique.php?id\_rubrique=49} 
package by E. Bertin.  These images were all obtained under photometric 
conditions and calibrated with multiple observations of standard star fields 
from the data compiled by 
P.B. Stetson\footnote{http://cadcwww.hia.nrc.ca/standards/} onto the 
Vega magnitude system.  

We calculated astrometric solutions for the images with the WCSTools package 
\citep{mink02}, cataloged the sources with the SExtractor package 
\citep{bertin96}, and matched these sources with the \chandra\ source catalog 
to create prioritized catalogs for multislit spectroscopy. 
The SExtractor detection parameters were approximately the default settings 
and the most relevant include the requirement that each source include 
a minimum of six pixels at least 3$\sigma$ brighter than the sky level. 
Typically there were several X-ray sources with obvious and bright 
visible-wavelength counterparts in each field. These matches 
were used to fine tune the registration of the X-ray catalog to match the 
WCS solution for the visible-wavelength catalogs. Sources were targeted for 
spectroscopy if they were within a generous $2''$ of the centroid of an X-ray 
source. Small $R-$band images of each X-ray source were also generated 
to inspect the quality of each match and the size of the X-ray PSF was 
considered in this inspection. 
The observed visible-wavelength photometry for all but three galaxies in our 
sample are provided in Table~\ref{tbl:catalog}. 
The photometric measurements were obtained from SExtractor BEST magnitudes, 
which for these relatively crowded fields are a corrected isophotal magnitude 
calculated after subtraction of neighboring galaxies that may contaminate 
the photometry. 
These three galaxies fell outside (or at the edge) of the field of view of 
our TEK5 or WFCCD images and are discussed further below. Visible-wavelength 
images of all of these sources are shown in Figure~\ref{fig:charts} along with 
the 95\% X-ray error circles. 
These error circles were calculated by generating fake sources with the 
MARX\footnote{http://space.mit.edu/CXC/MARX/} software and comparing their 
input positions to the measured positions with wavdetect. 
The distribution of the differences in position were used to 
calculate the 95\% X-ray error circle, where the size of the error circle 
depends on off-axis angle and the number of counts from the source. 
This technique will be described in more detail in M. Kim \etal\ (2006, {\it in prep}). 
In all but two cases (Abell~3125-6 and Abell~3128-5) these images 
shown in Figure~\ref{fig:charts} are from our $R-$band observations. Images 
of the two exceptions were obtained from the {\it Digitized Sky Survey}. 

  \subsection{Spectroscopy} 

\begin{figure*}
\figurenum{2} 
\plotone{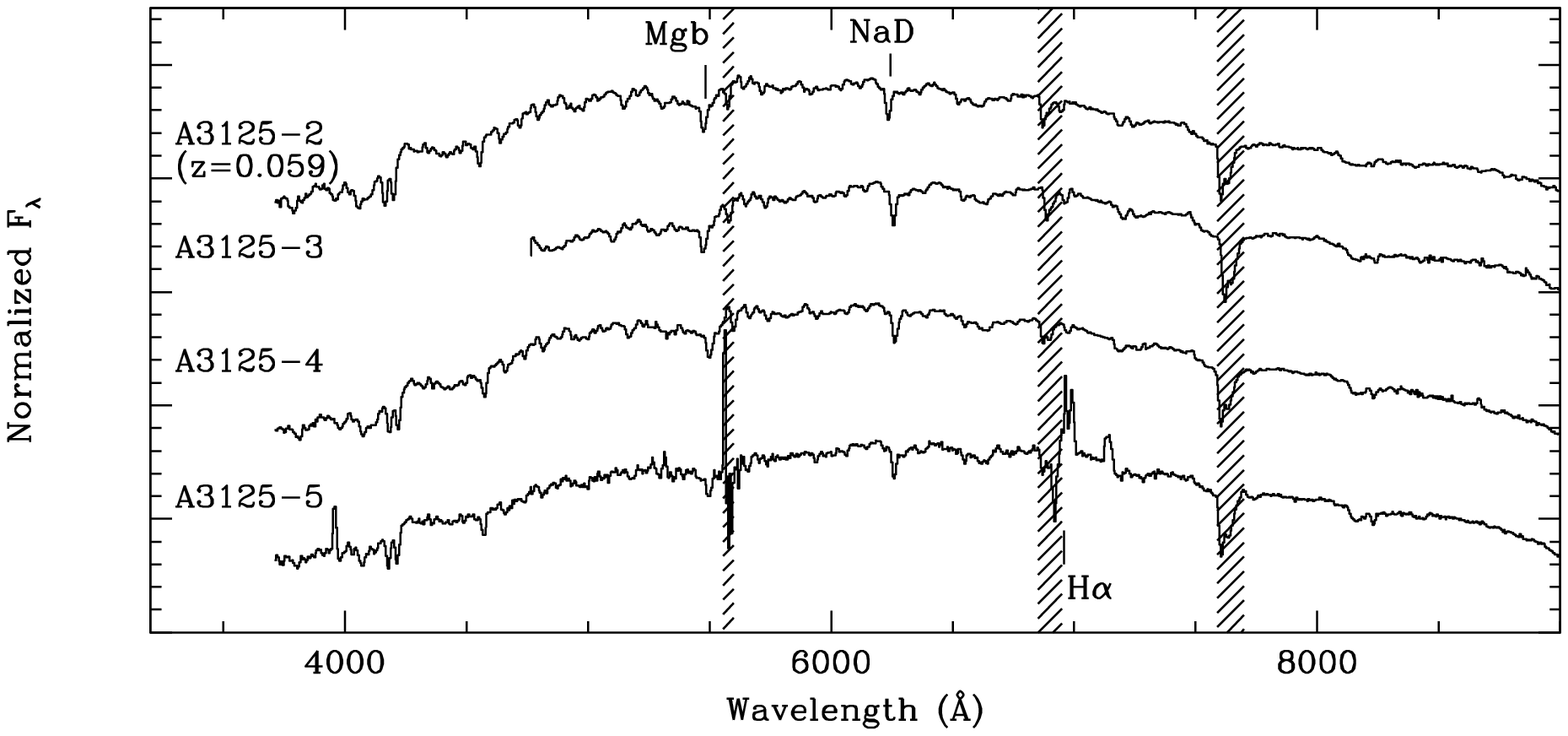} 
\caption{
Spectra of X-ray sources in Abell~3125, Abell~3128, Abell~644, Abell~1689, 
Abell~2163, MS1008, and AC~114. The hatched vertical lines 
correspond to bands of significant telluric absorption. Several prominent 
emission and absorption lines are labelled. The spectra are offset vertically 
from one another to minimize overlap between the objects.
\label{fig:spectra} 
}
\end{figure*}

\begin{figure*}
\figurenum{2} 
\plotone{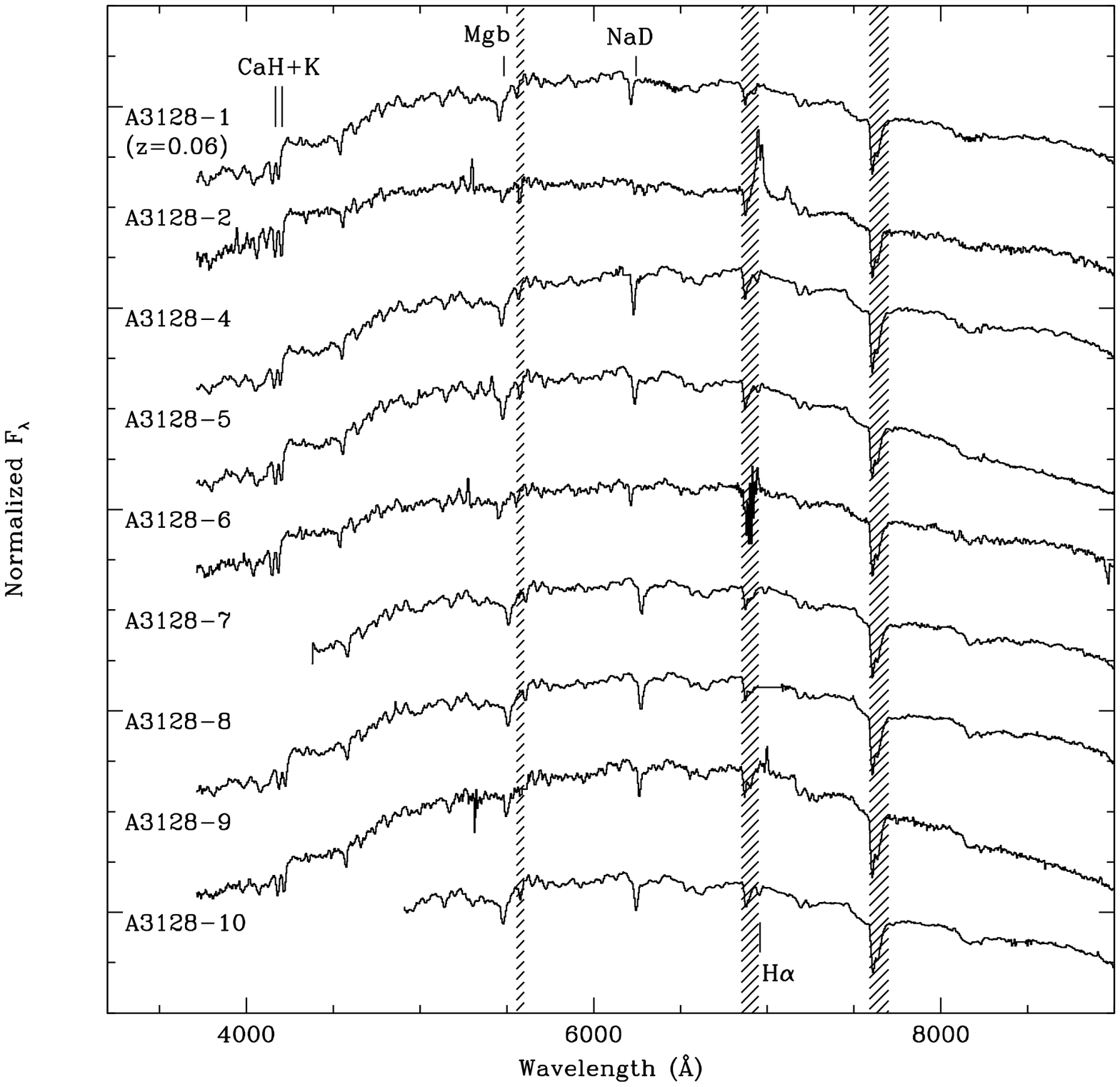} 
\caption{{\it Continued}}
\end{figure*}

\begin{figure*} 
\figurenum{2} 
\plotone{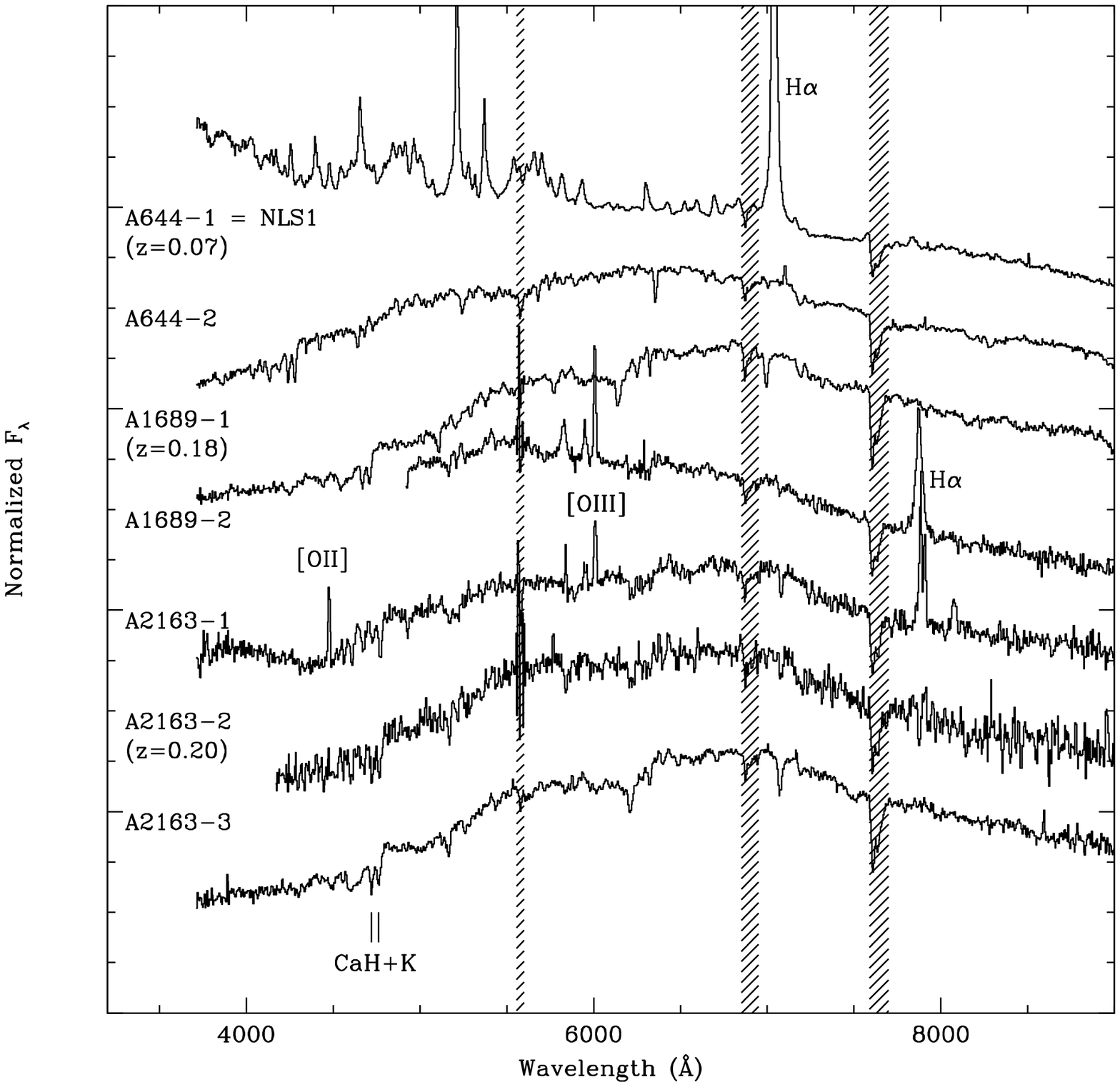} 
\caption{{\it Continued}}
\end{figure*}

\begin{figure*}
\figurenum{2} 
\plotone{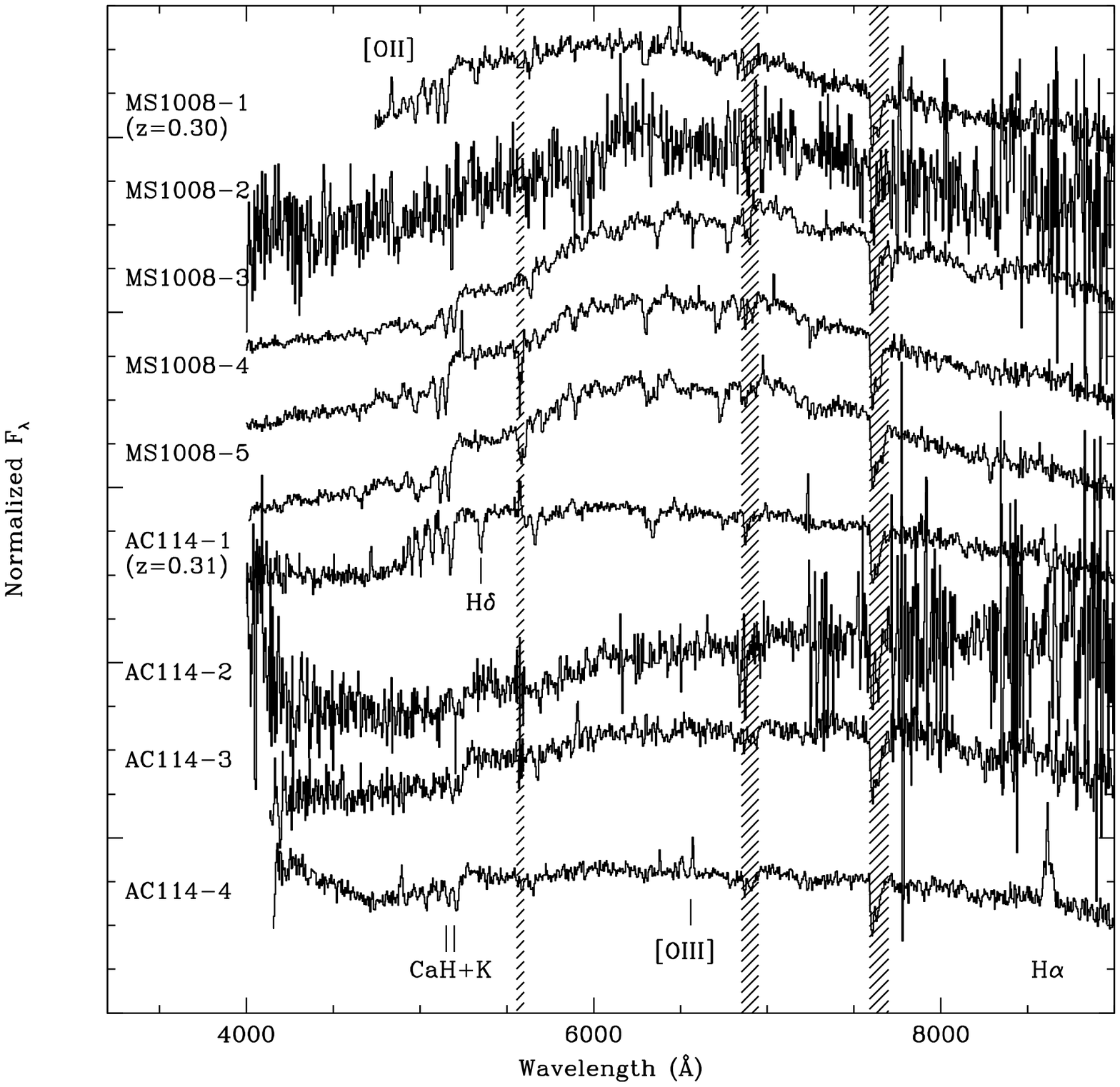} 
\caption{{\it Continued}}
\end{figure*}

The X-ray and visible-wavelength photometric catalogs were used to design 
multislit masks for spectroscopic observations with the LDSS-2 spectrograph 
on the 6.5m Clay telescope of the Magellan Project. 
All X-ray sources with $R-$band counterparts in our deep images from the 
du Pont telescope, corresponding to sources with $R < 24$ mag, were given 
first priority. 
Second priority was assigned to other bright, resolved sources. 
Each mask included approximately 15 -- 25 slits, where lower-redshift 
clusters typically had fewer sources per masks. 
We observed 3 -- 10 masks per cluster and 5 -- 10 X-ray sources per mask. 
All masks were designed with $1''$ slit widths. 
For some of the lower-redshift clusters we also designed additional masks 
without X-ray sources in order to obtain more data on the cluster galaxy 
population. 
All of the X-ray sources were assigned to at least one multislit observation. 
Each mask was observed for between 1200 and 5400s, with 
longer observations for higher-redshift clusters. Flatfields and 
comparison lamps were observed before and/or after each mask. 
We obtained multislit observations for seven clusters and present new 
photometry and analysis of our previously-published spectra of the 
six galaxies in Abell~2104 \citep{martini02}. 

The spectroscopic observations were processed with a series of custom Python 
scripts\footnote{Available at http://www.ociw.edu/ $\sim$kelson}.  
After a bias frame was subtracted from each raw data frame, the spatial 
and spectral distortions of each image were calculated from spectroscopic 
observations of flat field and HeNeAr comparison lamp observations, 
respectively, following the techniques described in \citet{kelson00}. 
The comparison lamps were then used to calculate a fifth-order wavelength 
solution for each frame. The zeropoint of the wavelength solution 
was then recalculated from prominent night sky lines in the red region of 
the spectrum. 
The flatfield observations were used to correct the frames and the flats 
themselves for variable slit illumination and normalized flatfields 
were used to removing the effects of fringing from the spectra. 
A model for the night sky emission was then fit to each frame following the 
method described in \citet{kelson03}. After this sky model was subtracted from 
each frame, the centroids and spatial extent of the sources were tabulated 
for all of the observations of a given mask. 
One-dimensional spectra for each object were extracted from all of the 
observations of a given mask simultaneously using the B-spline algorithm 
described in \citet{kelson06}. 
This method accounted for the different distortions and 
dispersion solutions in each frame and generated linear output spectra
with a dispersion of 5 \AA\ pix$^{-1}$. The spectra extend from 4000 
\AA\ to 9000 \AA, except for some that fall near an edge of the mask. 
The final output spectrum for each source includes the object counts, 
the sky counts, and the signal-to-noise per pixel. 

Redshifts for all of the sources were calculated using a cross-correlation 
technique in pixel space, rather than Fourier space, which has the advantage 
that each pixel may be weighted by the inverse of the variance, rather than 
assigned equal weight. This technique is more fully described in 
\citet{kelson05}. 
A number of our spectroscopic targets, both with and without X-ray emission, 
have previously published redshifts, including Abell~3125 and Abell~3128 
\citep{rose02}, MS1008 \citep{yee98}, and AC~114 \citep{couch01}. 
These literature data were used to cross-check our redshift measurements 
as well as to determine the redshift range of likely cluster members. 
For clusters without published velocity dispersions, we conservatively 
estimated membership if the galaxies were within 2000 \kms of the cluster mean 
velocity. Only one source (A1689-2) has a substantial offset of slightly 
over 4000 \kms\ from the cluster mean. This source is included in the 
cluster sample, although it is flagged with a $\Delta z$ in 
Table~\ref{tbl:catalog}. 

We have also flagged (with a $\Delta r$) seven sources whose visible-wavelength 
centroids are only marginally within the X-ray error circles. These sources 
could be either off-nuclear X-ray sources or chance superpositions of 
X-ray sources with a cluster member. The probability of a chance superposition 
of a cluster member with a background (or foreground) X-ray source depends 
strongly on the sensitivity of the \chandra\ dataset. From the X-ray log 
$N$ -- log $S$ relation, we expect approximately 1000 X-ray sources per 
square degree, or about one every three square arcminutes \citep{kim04b}. 
This is comparable to the average surface density of cluster members to 
our absolute magnitude limit and therefore chance superpositions must be 
extremely rare. A second possibility is that the X-ray sources with large 
$\Delta r$ are associated with the cluster galaxies, but they are ultraluminous 
X-ray sources rather than nuclear emission. In this case their high 
luminosity ($L_X \sim 10^{41}$ \ergs) makes this conclusion unlikely because 
known ultraluminous X-ray sources rarely have $L_X > 10^{40}$ \ergs\ 
\citep{swartz04}. 

The redshifts for the counterparts to all of the cluster X-ray sources are 
provided in Table~\ref{tbl:catalog}. 
The star formation properties of this sample can be studied with 
measurements of the \oii\ and \hd\ equivalent widths, where 
we have tabulated these values for our sample in Table~\ref{tbl:catalog} 
using the bandpass definitions from \citet{fisher98}. We detect \oii\ 
emission in five galaxies and \hd\ absorption in eight galaxies. Detection 
limits for the remaining sources vary, but typically are less than 2~\AA. 
Table~\ref{tbl:catalog} also notes any other prominent emission 
lines or an AGN classification if the emission-line ratios suggest an AGN 
based on the standard diagnostic diagram of \citet{baldwin81}. 
Spectra for these sources are shown in Figure~\ref{fig:spectra}, with the 
exception of the six sources in Abell~2104 published in \citet{martini02}. 

Table~\ref{tbl:lum} lists the absolute $R-$band magnitude $M_R$ for each 
galaxy. Corrections for bandpass shifting and stellar evolution have been 
applied based on a simple stellar population model with solar metallicity 
and formation redshift of $z = 3$ \citep{bruzual03}. 
This model was also used to correct the $B-R$ colors to the rest-frame values 
shown in the relevant figures. 
We estimated the spectroscopic completeness for these observations by 
calculating the fraction of spectroscopic candidates with redshifts as a 
function of $R-$band magnitude. The 95\% completeness varies between 
$R = 20 - 22$ mag and increases for higher-redshift clusters. For all 
of the clusters our redshift catalog is complete for all X-ray 
counterparts with rest-frame $M_R < -20$ mag. 

\begin{deluxetable*}{lccccccc}
\tablecolumns{8}
\tablenum{4}
\tabletypesize{\scriptsize}
\tablecaption{Fluxes and Luminosities\label{tbl:lum}}
\tablehead{
\colhead{ID} &
\colhead{$M_R$} & 
\colhead{$f_X$ (Broad)} & 
\colhead{$f_X$ (Soft)} & 
\colhead{$f_X$ (Hard)} & 
\colhead{$L_X$ (Broad)} & 
\colhead{$L_X$ (Soft)} & 
\colhead{$L_X$ (Hard)} \\
\colhead{(1)} &
\colhead{(2)} &
\colhead{(3)} &
\colhead{(4)} &
\colhead{(5)} &
\colhead{(6)} &
\colhead{(7)} &
\colhead{(8)} \\
\colhead{} &
\colhead{} &
\multicolumn{3}{c}{------------------ \ergs cm$^{-2}$ ------------------} & 
\multicolumn{3}{c}{------------------ \ergs ------------------} \\
}
\startdata 
A3125-1  & -21.19  & 2.6e-14 (7.0e-15) & 2.6e-14 (7.1e-15) &       ...         & 2.3e+41 & 2.3e+41 &    ...  \\
A3125-2  & -22.02  & 2.1e-14 (7.6e-15) & 2.3e-14 (7.6e-15) &       ...         & 1.9e+41 & 2.0e+41 &    ...  \\
A3125-3  & -21.86  & 9.5e-15 (5.1e-15) & 1.1e-14 (5.3e-15) &       ...         & 8.2e+40 & 9.6e+40 &    ...  \\
A3125-4  & -21.86  & 9.7e-15 (5.0e-15) & 7.9e-15 (4.8e-15) &       ...         & 8.4e+40 & 6.9e+40 &    ...  \\
A3125-5  & -21.04  & 3.4e-14 (8.3e-15) & 1.6e-14 (6.4e-15) & 9.4e-15 (3.3e-15) & 2.9e+41 & 1.4e+41 & 8.2e+40 \\
A3125-6  &   ...   & 2.4e-14 (9.3e-15) & 2.3e-14 (8.8e-15) &       ...         & 2.1e+41 & 2.0e+41 &    ...  \\
A3128-1  & -21.65  & 1.2e-14 (3.4e-15) & 1.1e-14 (3.5e-15) &       ...         & 1.0e+41 & 9.9e+40 &    ...  \\
A3128-2  & -19.96  & 1.1e-13 (9.1e-15) & 8.3e-14 (8.2e-15) & 1.7e-14 (2.7e-15) & 9.9e+41 & 7.5e+41 & 1.6e+41 \\
A3128-3  &   ...   & 1.0e-14 (4.1e-15) & 8.0e-15 (3.7e-15) & 1.4e-15 (1.4e-15) & 8.9e+40 & 7.2e+40 & 1.2e+40 \\
A3128-4  & -22.12  & 6.0e-14 (6.8e-15) & 6.2e-14 (7.0e-15) &       ...         & 5.4e+41 & 5.6e+41 &    ...  \\
A3128-5  &   ...   & 9.5e-15 (4.1e-15) & 7.0e-15 (3.6e-15) & 1.4e-15 (1.4e-15) & 8.5e+40 & 6.3e+40 & 1.2e+40 \\
A3128-6  & -20.20  & 1.2e-13 (8.4e-15) & 9.5e-14 (7.8e-15) & 1.1e-14 (2.1e-15) & 1.0e+42 & 8.5e+41 & 9.9e+40 \\
A3128-7  & -22.01  & 1.9e-14 (4.2e-15) & 2.0e-14 (4.2e-15) &       ...         & 1.7e+41 & 1.8e+41 &    ...  \\
A3128-8  & -21.85  & 1.5e-14 (4.7e-15) & 1.2e-14 (4.3e-15) & 1.9e-15 (1.3e-15) & 1.4e+41 & 1.0e+41 & 1.7e+40 \\
A3128-9  & -20.66  & 1.5e-14 (3.9e-15) & 1.2e-14 (3.5e-15) & 1.9e-15 (1.2e-15) & 1.4e+41 & 1.0e+41 & 1.7e+40 \\
A3128-10 & -22.22  & 2.5e-14 (5.0e-15) & 2.6e-14 (5.1e-15) &       ...         & 2.3e+41 & 2.3e+41 &    ...  \\
A644-1   & -20.61  & 1.0e-13 (6.9e-15) & 6.4e-14 (5.4e-15) & 1.7e-14 (2.1e-15) & 1.3e+42 & 8.0e+41 & 2.1e+41 \\
A644-2   & -21.58  & 1.3e-14 (3.4e-15) & 2.6e-15 (2.1e-15) & 5.2e-15 (1.5e-15) & 1.7e+41 & 3.3e+40 & 6.5e+40 \\
A2104-1  & -22.06  & 5.5e-13 (9.1e-15) & 1.1e-13 (2.6e-15) & 1.7e-13 (4.0e-15) & 4.0e+43 & 7.9e+42 & 1.2e+43 \\
A2104-2  & -20.71  & 1.1e-14 (1.4e-15) & 1.3e-15 (3.6e-16) & 4.9e-15 (7.4e-16) & 7.9e+41 & 9.4e+40 & 3.5e+41 \\
A2104-3  & -22.20  & 4.0e-15 (9.9e-16) & 1.7e-15 (4.0e-16) &       ...         & 2.9e+41 & 1.2e+41 &    ...  \\
A2104-4  & -19.76  & 4.0e-15 (1.2e-15) & 1.8e-15 (4.9e-16) &       ...         & 2.9e+41 & 1.3e+41 &    ...  \\
A2104-5  & -19.80  & 2.3e-14 (2.0e-15) & 7.9e-15 (7.5e-16) & 2.5e-15 (5.6e-16) & 1.7e+42 & 5.7e+41 & 1.8e+41 \\
A2104-6  & -21.59  & 4.0e-15 (1.0e-15) & 7.0e-16 (3.2e-16) & 1.4e-15 (4.6e-16) & 2.9e+41 & 5.1e+40 & 1.0e+41 \\
A1689-1  & -22.00  & 1.2e-14 (5.3e-15) & 1.3e-14 (5.8e-15) &       ...         & 1.2e+42 & 1.3e+42 &    ...  \\
A1689-2  & -21.41  & 1.8e-14 (6.2e-15) & 1.4e-14 (6.1e-15) & 2.5e-15 (2.0e-15) & 1.9e+42 & 1.5e+42 & 2.7e+41 \\
A2163-1  & -22.24  & 2.8e-14 (2.5e-15) & 1.8e-15 (9.1e-16) & 1.3e-14 (1.2e-15) & 3.8e+42 & 2.4e+41 & 1.7e+42 \\
A2163-2  & -19.96  & 5.6e-15 (1.2e-15) & 3.5e-15 (8.4e-16) & 8.3e-16 (3.7e-16) & 7.6e+41 & 4.8e+41 & 1.1e+41 \\
A2163-3  & -21.66  & 2.0e-15 (9.0e-16) & 1.8e-15 (7.1e-16) &       ...         & 2.7e+41 & 2.4e+41 &    ...  \\
MS1008-1 & -21.35  & 3.3e-15 (1.7e-15) & 1.7e-15 (1.3e-15) & 7.5e-16 (6.9e-16) & 1.2e+42 & 5.9e+41 & 2.6e+41 \\
MS1008-2 & -19.42  & 2.0e-15 (1.0e-15) & 1.2e-15 (8.1e-16) &       ...         & 7.0e+41 & 4.2e+41 &    ...  \\
MS1008-3 & -21.78  & 1.8e-14 (2.6e-15) & 1.2e-14 (2.0e-15) & 2.1e-15 (7.4e-16) & 6.3e+42 & 4.2e+42 & 7.3e+41 \\
MS1008-4 & -21.24  & 2.2e-15 (1.2e-15) & 1.6e-15 (9.6e-16) &       ...         & 7.7e+41 & 5.6e+41 &    ...  \\
MS1008-5 & -21.99  & 9.7e-15 (2.9e-15) & 6.6e-15 (2.3e-15) & 1.1e-15 (8.0e-16) & 3.4e+42 & 2.3e+42 & 3.8e+41 \\
AC114-1  & -21.03  & 8.1e-14 (2.8e-15) & 4.3e-14 (1.5e-15) & 1.2e-15 (3.6e-16) & 3.1e+43 & 1.7e+43 & 4.6e+41 \\
AC114-2  & -19.62  & 4.1e-15 (9.6e-16) & 1.5e-15 (4.6e-16) & 8.5e-16 (3.3e-16) & 1.6e+42 & 5.6e+41 & 3.2e+41 \\
AC114-3  & -21.78  & 1.1e-14 (1.1e-15) & 5.2e-15 (5.8e-16) & 7.8e-16 (2.9e-16) & 4.2e+42 & 1.9e+42 & 3.0e+41 \\
AC114-4  & -20.06  & 5.4e-15 (8.3e-16) & 1.1e-15 (3.1e-16) & 2.2e-15 (4.2e-16) & 2.0e+42 & 4.2e+41 & 8.4e+41 \\
AC114-5  & -20.02  & 7.7e-16 (5.2e-16) & 3.1e-16 (2.6e-16) &       ...         & 2.9e+41 & 1.2e+41 &    ...  \\
AC114-6  & -22.90  & 6.7e-16 (5.3e-16) & 3.5e-16 (2.7e-16) &       ...         & 2.5e+41 & 1.3e+41 &    ...  \\
\enddata
\tablecomments{
Absolute, rest-frame magnitude $M_R$, observed X-ray fluxes, and rest-frame 
X-ray luminosities. The X-ray fluxes and luminosities were determined with 
a $\Gamma = 1.7$ power law and Galactic absorption. 
}
\end{deluxetable*}

\section{Results} \label{sec:res} 

We have identified a total of 40 galaxies with X-ray emission in eight 
clusters, corresponding to between two and ten X-ray sources per cluster. 
Thirty four of these cluster members are new to this study, while six were 
published in our previous study of Abell~2104. 
Of these 34 new galaxies, we identified 29 with our spectroscopic observations 
and five additional sources from cross-correlating our X-ray sources with 
redshift data from the literature. References to these sources are 
provided in the caption to Table~\ref{tbl:catalog}. 

As noted above, several of these sources were previously known to be cluster 
members, although only the \citet{smith03} study of A3128 previously studied 
the X-ray emission.
He noted that our sources A3128-2 and A3128-6 had hard X-ray spectra and may 
be AGN, while of the three other galaxies A3128-7 and A3128-10 appeared to be 
normal E/SOs and A3128-6 was an E/SO with some shell structure. 
In the next sections below we describe the photometric and spectroscopic 
properties of this sample of cluster galaxies with X-ray emission. 
We then use these results to identify the likely physical mechanisms 
responsible for the X-ray emission and calculate the fraction of cluster 
galaxies that host AGN. 

\subsection{Visible-wavelength data} 

\begin{figure*}
\figurenum{3} 
\plotone{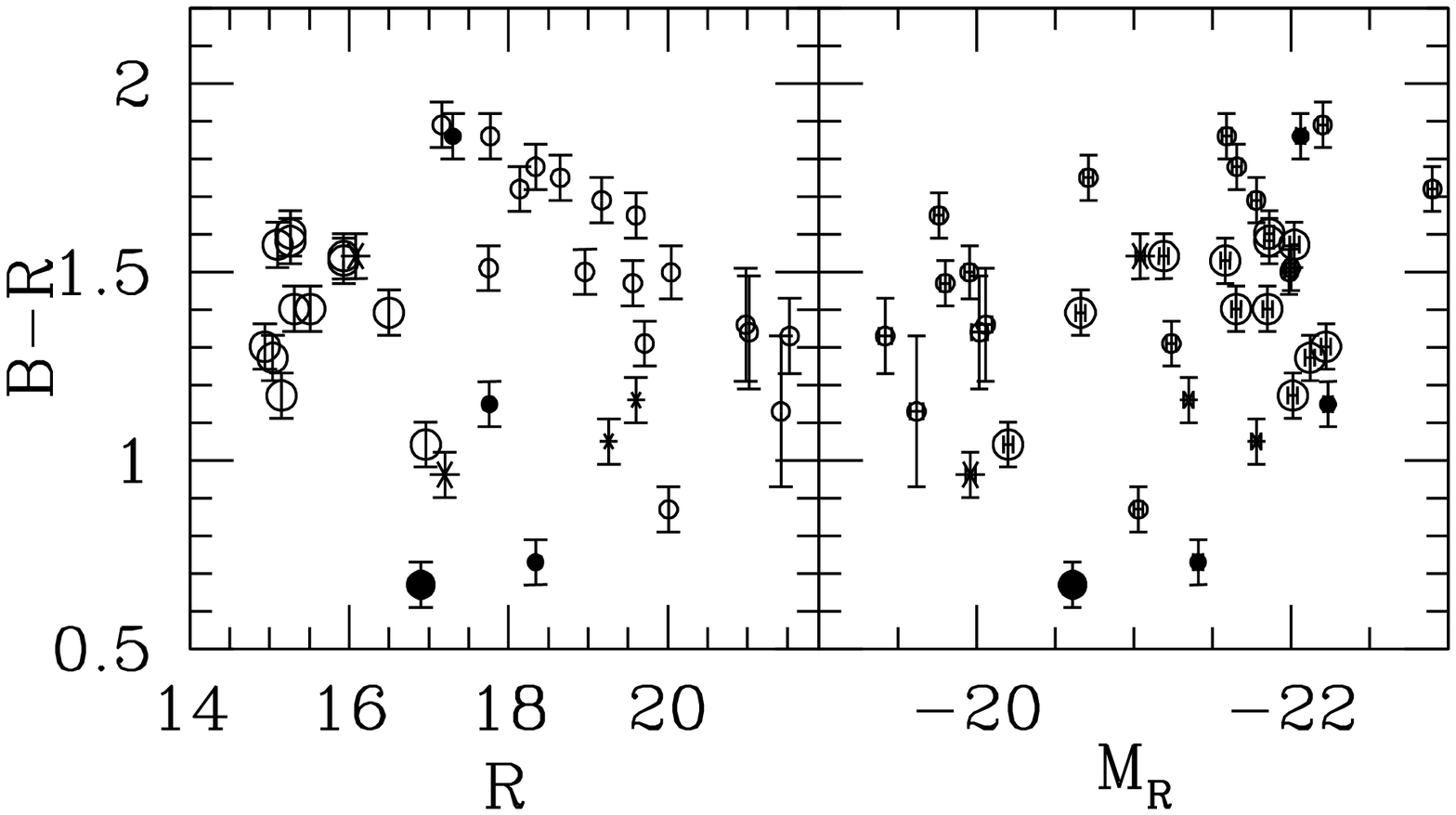} 
\caption{
$B-R$ vs.\ $R$ ({\it left}) and vs.\ $M_R$ ({\it right}) for the cluster 
X-ray sources. Open circles mark cluster members without emission lines, 
asterisks are those with \oii\ emission, and filled circles indicate galaxies 
with AGN spectral signatures.  Galaxies with $z<0.1$ are shown with 
large symbols and $z>0.1$ with small symbols. 
The error bars are $1 \sigma$ in all of the plots.
\label{fig:brrmr} 
}
\end{figure*}

The $R-$band images of these galaxies (Figure~\ref{fig:charts}) show that 
most appear to be early-type galaxies, consistent with the morphological type 
most commonly found at the centers of rich clusters. 
This is particularly true of the lowest-redshift ($z < 0.1$) clusters, 
where all galaxies except for A3125-5 appear to be Es or S0s. Six of the 40 
galaxies show obvious signs of interactions. 

The left panel of Figure~\ref{fig:brrmr} presents a color-magnitude diagram of 
$B-R$ vs.\ $R$  and demonstrates that the X-ray counterparts with emission 
lines, particularly those with obvious AGN spectra signatures, tend to be 
bluer than galaxies that do not exhibit bright emission lines. This 
result is more pronounced when all of the lower-redshift ($z < 0.1$; large 
symbols) galaxies are compared separately from the higher-redshift 
($z > 0.1$) galaxies. Two of the four galaxies with AGN spectral signatures 
are the bluest galaxies in the sample. 
Sources with other strong emission lines, such as \ha, \oii, or \oiii\  
also tend to be substantially bluer than the 
other galaxies. In the lower-redshift subsample the emission-line galaxies 
are among the fainter sources, while in the higher-redshift subsample the 
obvious AGN and other emission line galaxies tend to be in the brighter 
half of the distribution. 
This is shown in right panel of Figure~\ref{fig:brrmr}, which plots $B-R$ 
vs.\ $M_R$. This figure demonstrates that the emission-line galaxies do 
not tend to reside in the most luminous host galaxies, but rather in 
galaxies only slightly more luminous than the knee of the galaxy luminosity 
function $M_R^* = -21.15$ mag \citep{christlein03}. 

\begin{figure}
\figurenum{4} 
\plotone{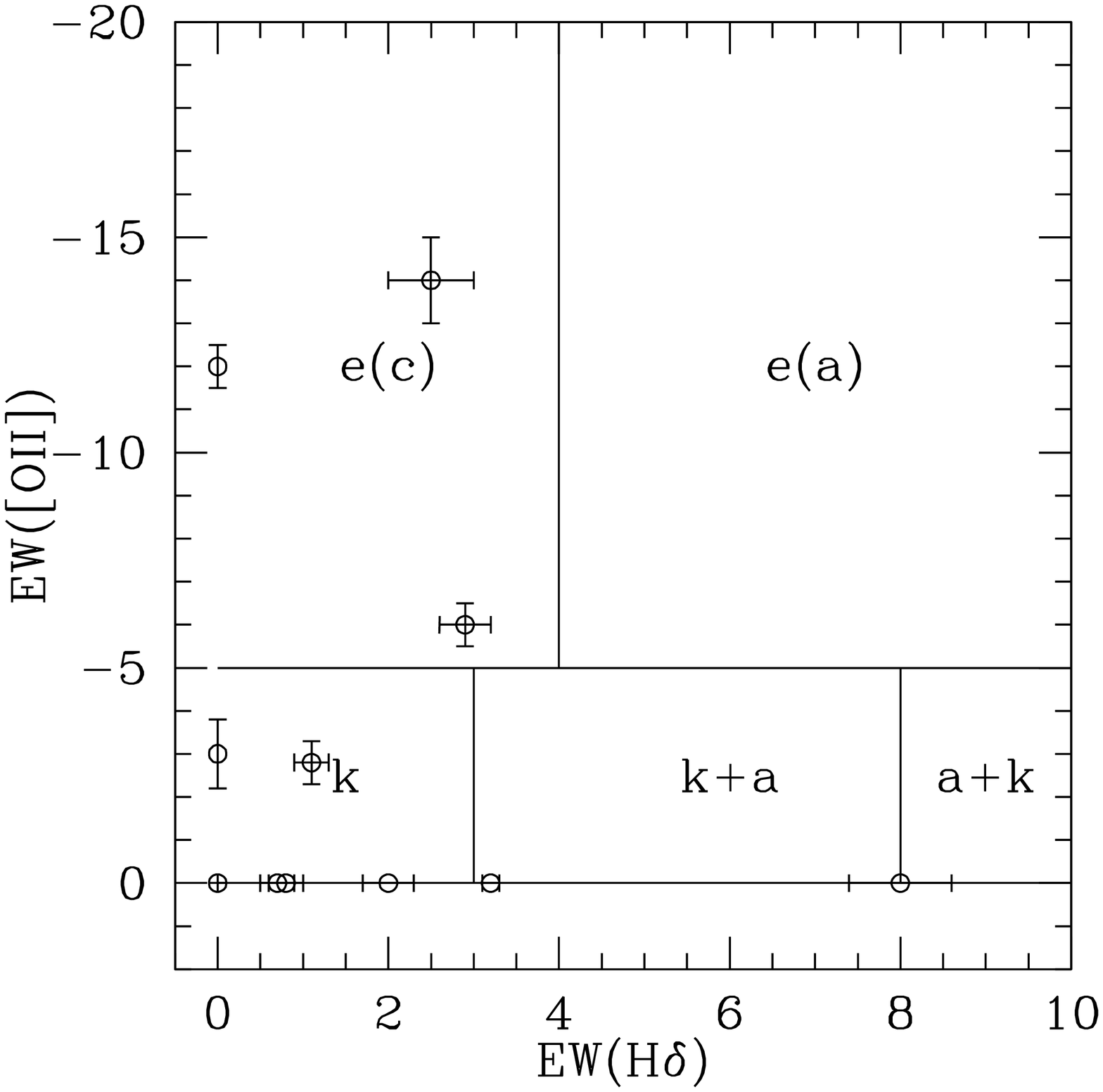} 
\caption{
Plot of [OII] vs.\ H$\delta$ equivalent width on a grid of the spectral 
classification scheme of \citet{dressler99}. We do not detect either of 
these lines from 20 of the 31 galaxies for which we have sufficient 
wavelength coverage. 
\label{fig:o2hd} 
}
\end{figure}

We use the \oii\ and \hd\ equivalent widths as estimators of the star formation 
history of these X-ray sources, where the \oii\ equivalent width is a measure 
of current star formation and \hd\ is a measure of the remnant population 
of a significant starburst in the recent past. 
Figure~\ref{fig:o2hd} illustrates the location of our sample galaxies in this 
parameter space, along with the spectral classification scheme developed by 
\citet{dressler99}. 
Only three of these galaxies are classified as e(c) and have \oii\ widths
greater than 5 \AA\ and moderate 
Balmer absorption. Two galaxies have no detected \oii, but \hd\ 
greater than 3 \AA\ indicative of moderate to strong Balmer absorption. 
The most extreme case is AC114-1, whose spectrum exhibits a strong Balmer 
series and is a borderline poststarburst galaxy. Most of the galaxies 
in this sample show no strong \oii\ or \hd. 

\subsection{X-ray Colors and Quantiles} 

X-ray colors offer a simple means of quantifying the properties of 
X-ray sources. 
As the X-ray colors are calculated directly from count rates, the 
correspondence between colors and spectral properties, such as power-law 
shape, depend on the energy sensitivity of a given detector. These differences 
are particularly acute between the front-illuminated (FI) ACIS-I 
(CCDs 0 through 3) and the back-illuminated (BI) ACIS-S (CCDs 5 and 7) and 
consequently we have calculated the conversion between X-ray colors and 
spectral properties for the ACIS-I and ACIS-S observations separately. 
The primary cause of this difference is that the BI chips are more sensitive 
to very soft X-ray photons ($< 1$ keV) and therefore a given source will, for  
example, register a larger $S_1$ count rate if observed with a BI chip 
relative to an FI chip. 

\begin{figure*}
\figurenum{5} 
\plotone{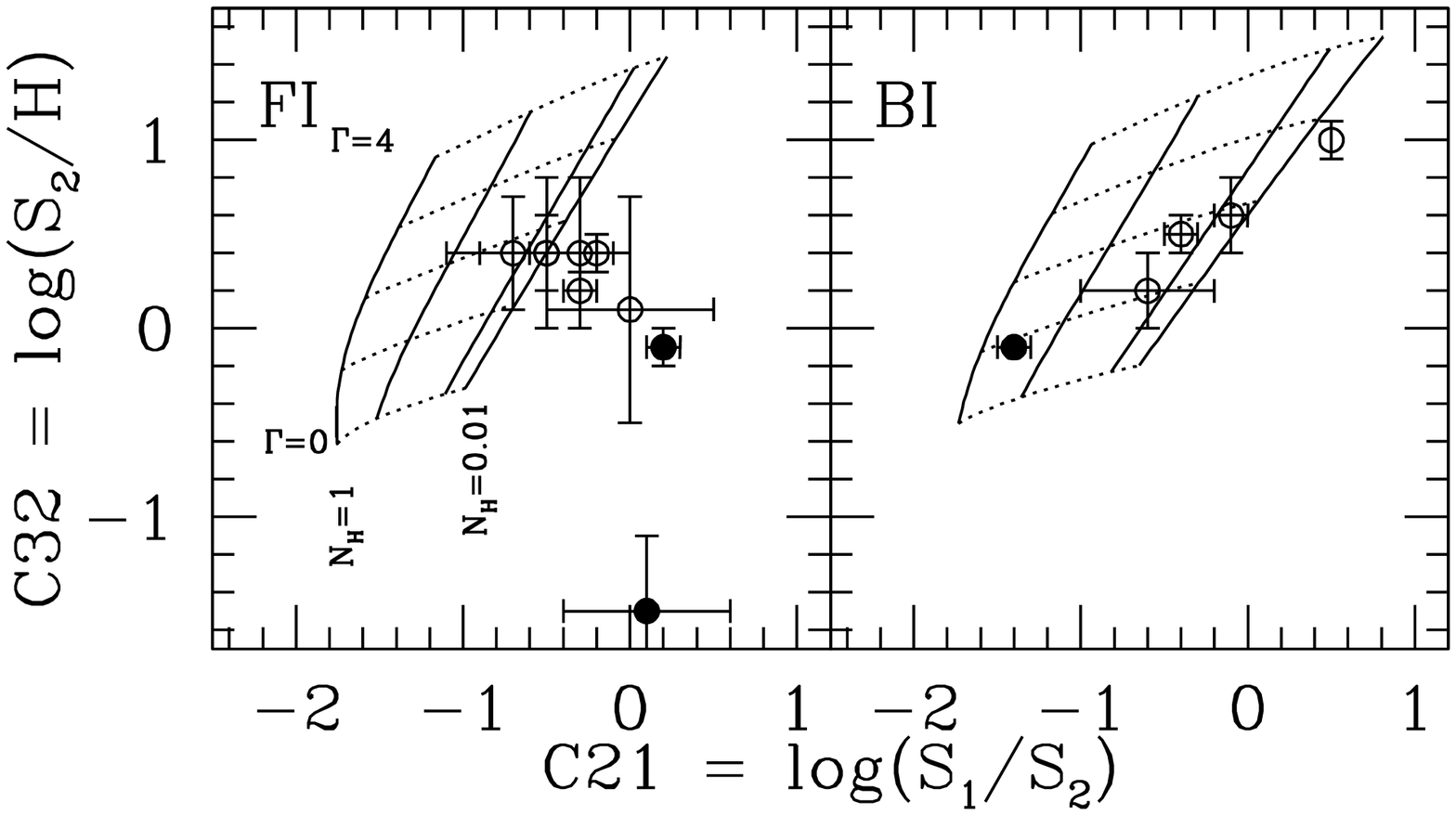} 
\caption{
X-ray colors C32 vs.\ C21. All sources with greater than three counts 
in the $S_1$, $S_2$, and $H$ bands are shown, along with grids representing 
the mean sensitivity of the observations with the the front-illuminated 
(left panel) and back-illuminated (right panel) ACIS chips. 
Spectroscopically-identified AGN are drawn with filled circles. 
The grids shown correspond to power-law models at $z=0$ with 
$\Gamma = 0, 1, 2, 3, 4$ (dotted lines) absorbed by 
$N_H = 0.01, 0.1, 0.5, 1 \times 10^{22}$ cm$^{-2}$. 
\label{fig:xccd} 
}
\end{figure*}

Figure~\ref{fig:xccd} displays the X-ray colors C32 vs.\ C21 for the 
FI and BI chips, along with grids that demonstrate the location of an 
absorbed power-law with $\Gamma = 0 - 4$ and $N_H = 0.01 - 1 \times 10^{22}$ 
cm$^{-2}$. The different shapes of the grids in the two panels reflects 
the substantial differences in the energy sensitivity of the FI and BI 
CCDs. The relative energy sensitivity of the chips change slightly with 
time and the grids shown are based on the average sensitivity of these 
observations. Sources are only shown if they have at least three counts in the 
$S_1$, $S_2$, and $H$ bands (see Table~\ref{tbl:xdata}). 
Nearly all of the sources shown in the figure are consistent with 
$\Gamma \sim 2$ and at most modest ($N_H \leq 10^{21}$ cm$^{-2}$) absorption. 
Those that do not include the spectroscopically-identified AGN 
({\it filled circles}) and indicate that an absorbed power-law is too 
simplistic a model for some of these sources. An absorbed power-law 
combined with emission from hot gas may be a better fit to these cases, 
although we lack sufficient signal-to-noise for detailed spectral 
modeling. 

\begin{figure*}
\figurenum{6} 
\plotone{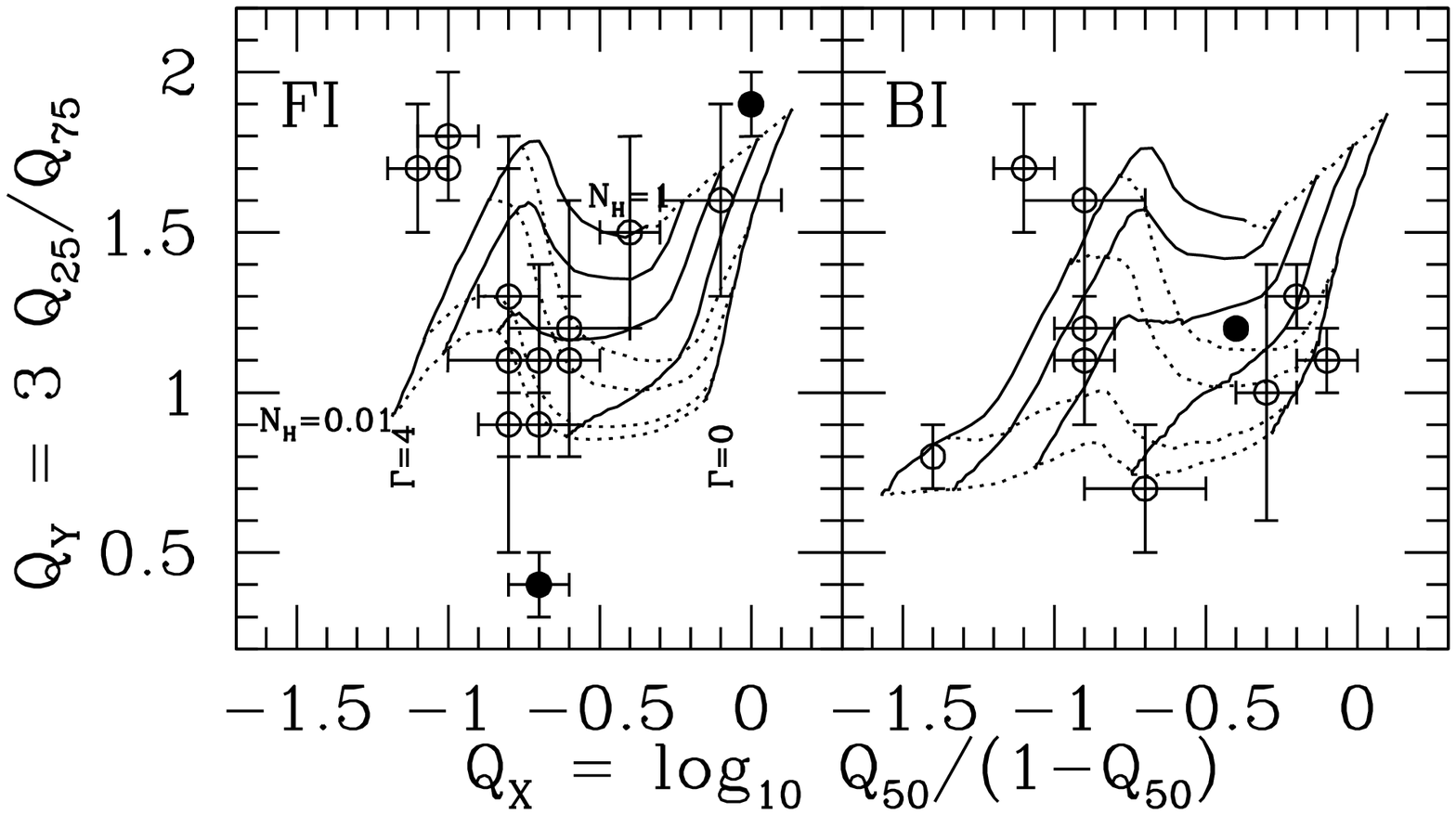} 
\caption{
X-ray quantiles. All sources with more than 25 broad band counts are 
shown, along with grids representing the mean sensitivity of the observations 
with the front-illuminated (left panel) and back-illuminated (right panel) 
ACIS chips. 
Spectroscopically-identified AGN are drawn with filled circles. 
The grids shown correspond to power-law models at $z=0$ with
$\Gamma = 0, 1, 2, 3, 4$ (dotted lines) absorbed by
$N_H = 0.01, 0.1, 0.5, 1 \times 10^{22}$ cm$^{-2}$.
\label{fig:xq} 
}
\end{figure*}

In a recent paper, \citet{hong04} argued that this traditional approach of 
using X-ray colors in predetermined bands is strongly biased toward 
particular values of $\Gamma$ and $N_H$ (for an assumed power-law model).
These authors show that the bands used to define the colors used in 
Figure~\ref{fig:xccd} can classify sources with $\Gamma \sim 2$ and 
modest absorption with only a few counts, while there will be substantial 
uncertainties in the classification of substantially softer, harder, and/or 
more obscured sources. Effectively, a source with higher or lower $\Gamma$ 
must be detected with substantially higher counts in order to fall in 
much of the traditional X-ray color space. The common adoption of an 
unobscured $\Gamma = 1.7$ power-law for typical X-ray sources may therefore 
be the result of selection effects due to the band definitions, rather than 
a reflection of the ubiquity and appropriateness of this model for faint X-ray 
sources. In Figure~\ref{fig:xq} we plot the X-ray quantiles defined in 
Section~\ref{sec:xdata} along with the same grids of $\Gamma$ and $N_H$ shown 
in Figure~\ref{fig:xccd} (the approximate axes of $\Gamma$ and $N_H$ are 
rotated by $\sim 90$ degrees). As in Figure~\ref{fig:xccd}, most of the 
sources are concentrated at low $N_H$ and $\Gamma \sim 2$, although there 
are many more very soft sources (large $\Gamma$) and several more hard sources 
than is apparent from the X-ray colors. As in the previous figure, the most 
significant outliers include the spectroscopically-confirmed AGN. This 
confirms our conclusion based on Figure~\ref{fig:xccd} that a simple, 
absorbed power-law model is not a good fit to even the obvious low-luminosity 
AGN, although more counts would be required to explore a wider 
parameter space of models. 

\subsection{Visible to X-ray Properties} 

\begin{figure}
\figurenum{7} 
\plotone{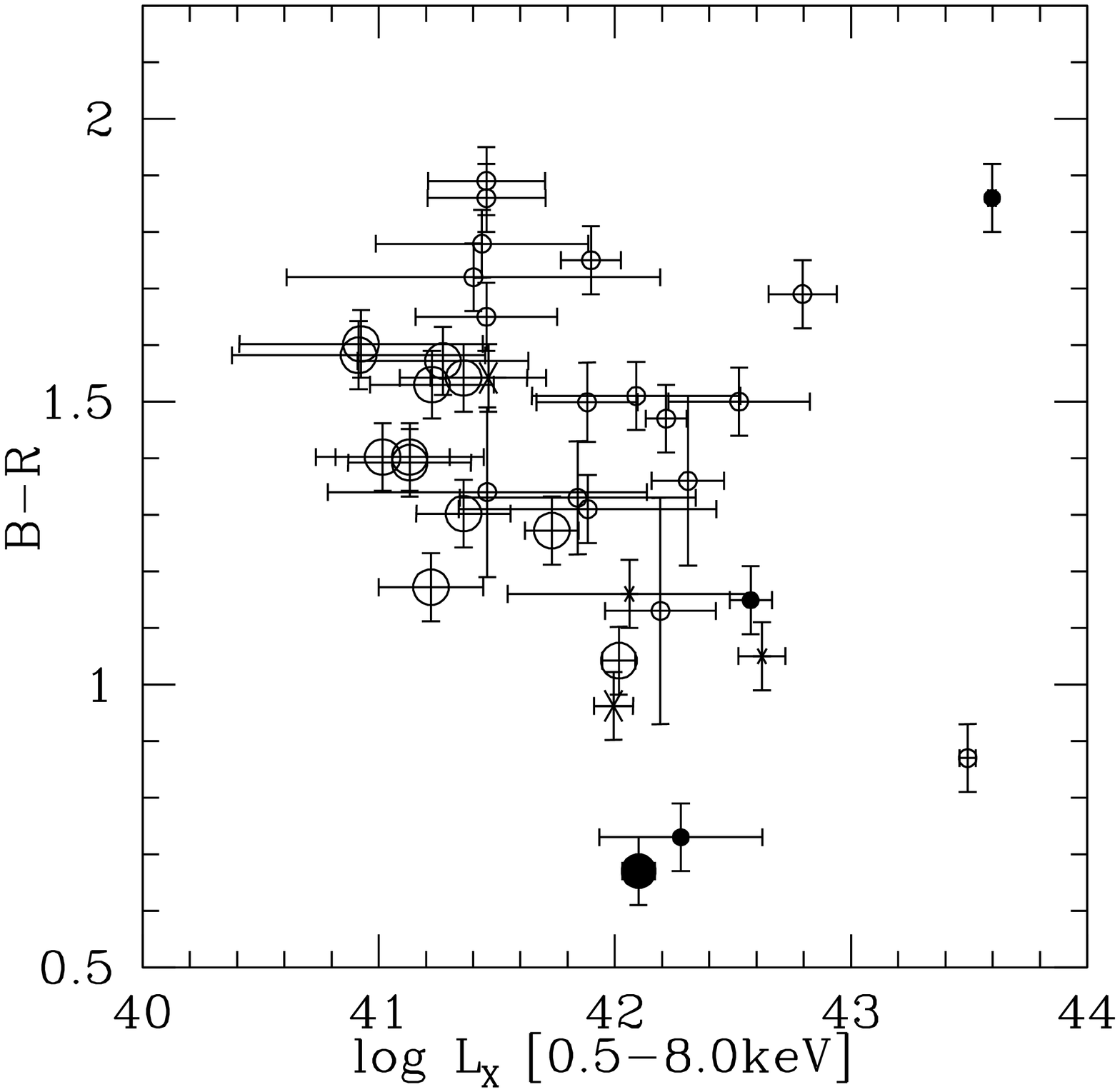} 
\caption{
$B-R$ vs.\ $L_{X,B}$ for the cluster X-ray sources. Symbols are as defined in 
Figure~\ref{fig:brrmr}. 
The error bars are $1 \sigma$ in all of the plots. 
\label{fig:brlxb} 
}
\end{figure}

Taken separately, the previous two sections present an apparent contradiction: 
From the visible-wavelength data alone, it appears that only four of these 
40 X-ray sources are AGN and the remainder are galaxies with modest 
star formation or passively-evolving galaxies. 
In contrast, the X-ray data indicate that most of 
these sources are quite luminous in X-rays and many are consistent with the 
unobscured, $\Gamma=1.7$ power laws typical of AGN. 
Furthermore, the spectroscopically-identified AGN are those least consistent 
with unobscured, $\Gamma=1.7$ power law emission. 
Here we investigate the visible to X-ray properties of these sources in 
order to resolve this contradiction. 

If the X-ray emission from these galaxies is predominantly due to AGN 
and the X-ray to visible-wavelength flux ratio is constant for AGN, 
then the galaxies with AGN spectral signatures and bluer colors should be 
among the most X-ray luminous sources. 
We test this hypothesis in Figure~\ref{fig:brlxb} with a plot of 
$B-R$ vs.\ broad band X-ray luminosity. This figure shows that only 
one\footnote{This is A2104-1, which was determined to be highly obscured by 
\citet{martini02}. The second most X-ray luminous galaxy is AC114-1, which has 
a post-starburst spectrum.} spectroscopically identifiable AGN is among the 
most X-ray luminous galaxies.  
Most of the X-ray sources with $L_X > 10^{42}$ \ergs\ do not show AGN spectral 
signatures, although we note that all of these galaxies are above $z=0.1$ 
and such signatures are more easily diluted by host galaxy starlight. 
These galaxies are generally not significantly bluer than other galaxies in 
these  clusters, although some would be classified as Butcher-Oemler galaxies 
\citep{butcher78}. 

\begin{figure*}
\figurenum{8} 
\plotone{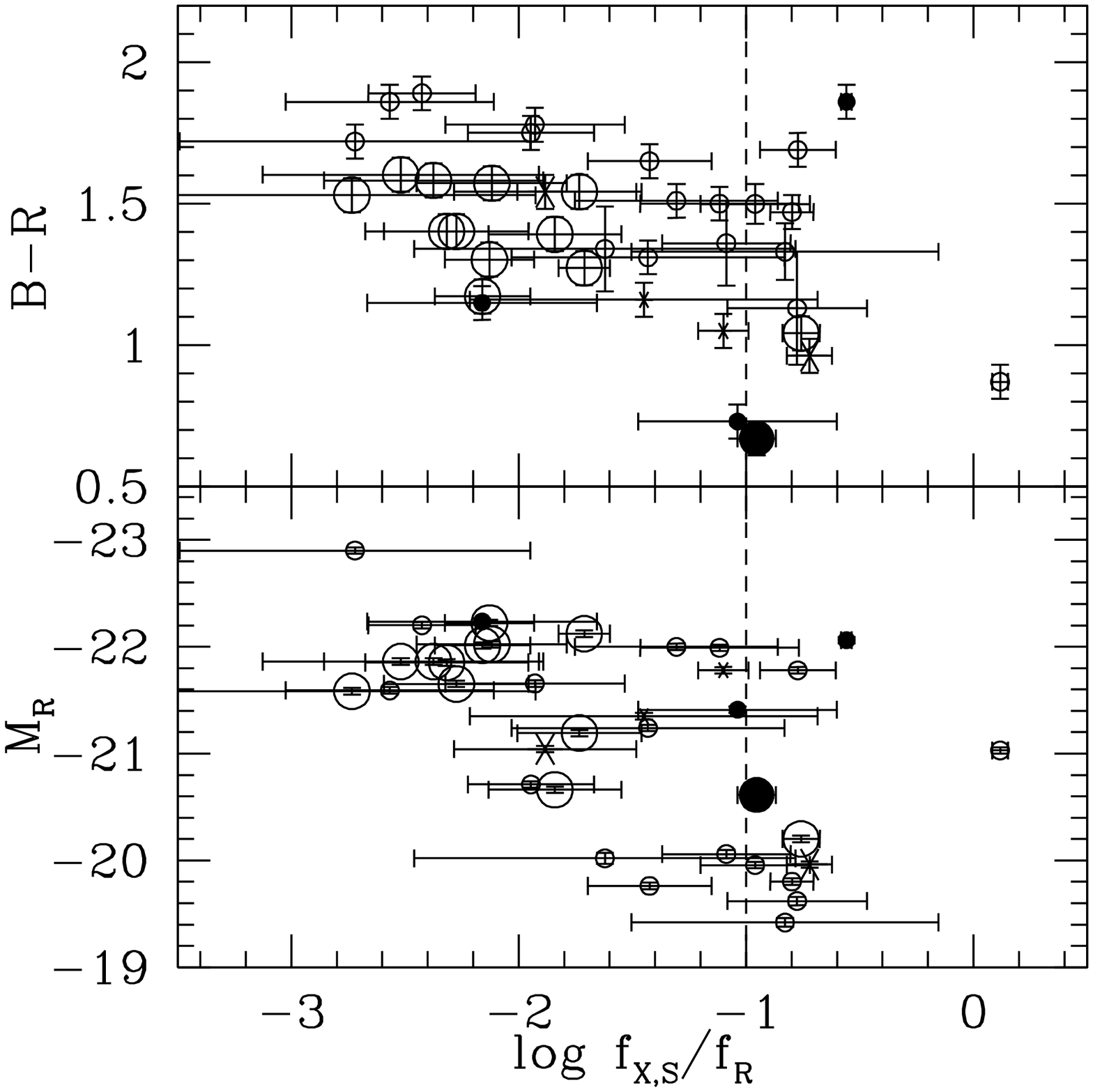} 
\caption{
$B-R$ (upper panel) and $M_R$ (lower panel) vs.\ $f_{X,S}/f_R$ for the 
cluster X-ray sources. Symbols are as defined in Figure~\ref{fig:brrmr}. 
\label{fig:brmrfxfr} 
}
\end{figure*}

The visible -- X-ray flux ratio can be another indicator of the presence 
of an AGN, as opposed to LMXBs or hot gas, because AGN have higher $L_X/L_B$ 
compared to other galaxies. 
In Figure~\ref{fig:brmrfxfr} we plot $B-R$ and $M_R$ vs.\ the soft X-ray to 
$R-$band flux ratio. 
This figure is similar to Figure~8 of \citet{kim04b}, which those authors 
use to separate various types of X-ray emitting galaxies in the CHaMP survey 
\citep[see also][]{green04}. 
The region to the right of the dashed vertical line at $f_X = 0.1 f_R$ and 
bluer than $B-R = 1$ is their locus for blue quasars. Sources with redder 
$B-R$ colors still have excess X-ray emission over that expected for hot 
X-ray halos or LMXBs and may therefore be obscured AGN. 
The very red A2104-1 is an excellent example of an obscured AGN with red 
visible-wavelength colors and excess X-ray emission. The other extreme 
source in this figure (at $B-R \sim 0.9$, $f_{X,S}/f_R \sim 0.1$) is the 
poststarburst galaxy AC114-1, which has an extremely soft X-ray spectrum.
Less than a third of our sample is located in their quasar regime 
defined by $f_X = 0.1 f_R$ and $B-R < 1$ in Figure~\ref{fig:brmrfxfr}, 
although this is not surprising as none of these X-ray sources have quasar 
luminosities ($L_X > 10^{44}$ \ergs) and therefore the $R-$band flux 
could include a contribution from stars. In fact nearly all of our 
spectra are dominated by host galaxy starlight. 
We investigate the $f_X/f_R$ ratio as a function of $M_R$ in the lower panel 
of Figure~\ref{fig:brmrfxfr} in order to determine the typical host galaxy 
luminosity of sources with large $f_X/f_R$. This figure demonstrates that 
most of the sources with large $f_X/f_R$ are in approximately 
$\sim M_R^*$ or fainter galaxies and are above $z = 0.1$.
The absence of low-luminosity galaxies with small $f_X/f_R$ is due to 
the sensitivity limit of the X-ray observations. While our spectroscopic 
observations are complete to $M_R \sim -20$ mag, the X-ray observations are 
only sensitive to galaxies with small $f_X/f_R$ in these clusters if they 
have luminous host galaxies. 
In contrast, the increasing number of luminous galaxies with lower $f_X/f_R$ 
is significant. For luminous galaxies this illustrates that they are more 
likely to host a faint X-ray source than a bright one. 

\begin{figure}
\figurenum{9} 
\plotone{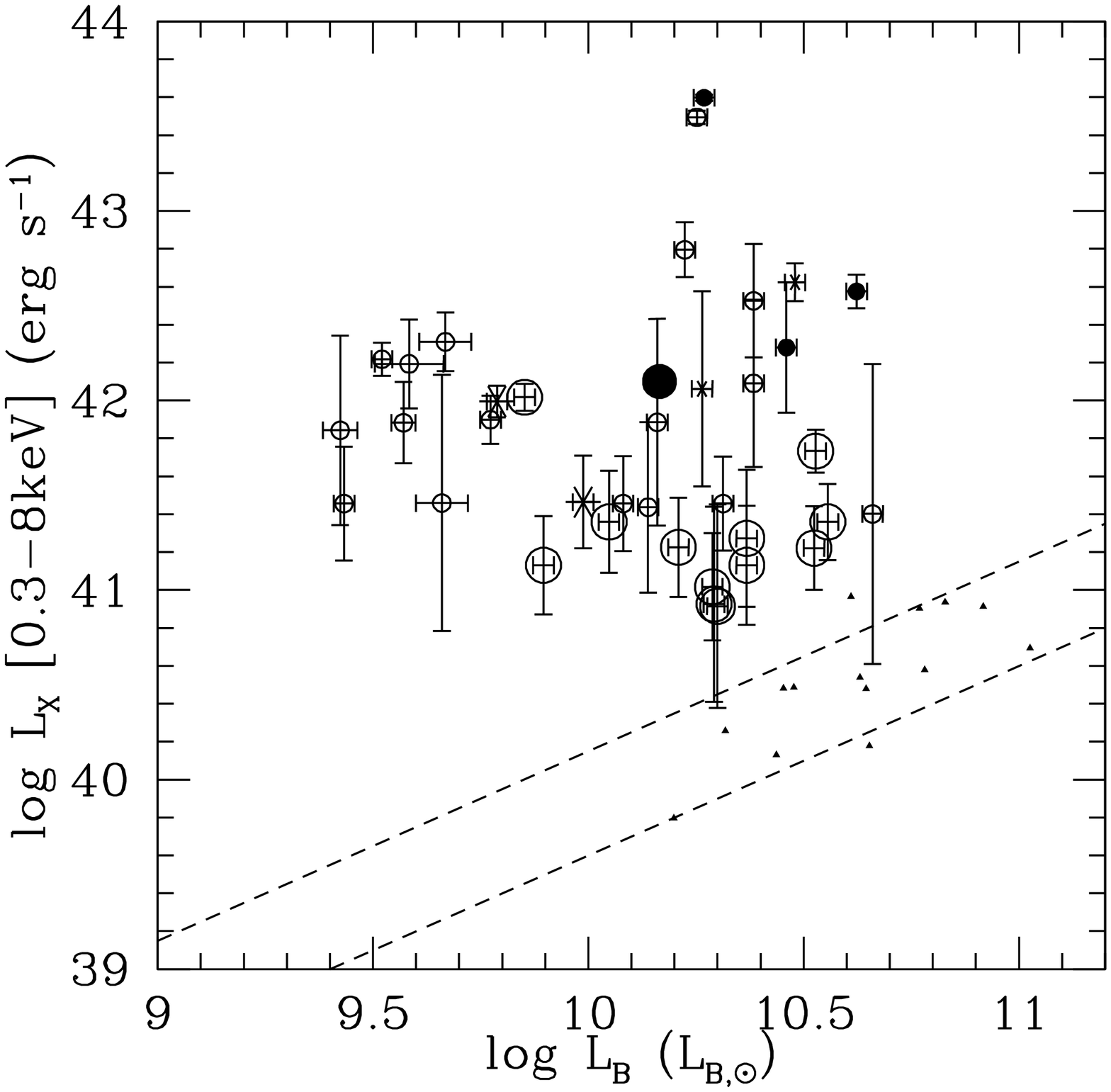} 
\caption{
$L_X$ [0.3--8 keV] vs.\ $L_B$ for all of our sources with $B-$band photometry. 
Most of these cluster sources lie well above the $L_X - L_B$ relation 
for X-ray emission due to LMXBs alone. The dashed lines bound the range 
defined by the standard deviation of the ratio quoted by \citet{kim04c}, 
while the small, filled triangles are the observations from that paper. 
\label{fig:lxlb} 
}
\end{figure}

Unlike the case for low-luminosity AGN, there is a relatively tight relation 
between X-ray and visible-wavelength flux for galaxies with X-ray emission 
due to LMXBs \citep{canizares87}. 
This relation has been studied in detail with \chandra\ observations of 
nearby elliptical galaxies \citep[e.g.][]{sarazin01,blanton01,helsdon01,
jeltema03,gilfanov04}. 
In a recent paper \citet{kim04c} derived the relation 
$L_X/L_B = 0.9 \times 10^{30}$ \ergs$/L_{B,\odot}$ based on broad-band X-ray 
measurements of 14 luminous, nearby galaxies. 
Figure~\ref{fig:lxlb} shows the broad band $L_X$ vs.\ $L_B$ for our 
sample, along with the galaxies measured by \citet{kim04c} and the band that 
defines the $1\sigma$ scatter in the $L_X/L_B$ ratio for LMXBs. 
This figure shows that all of these galaxies are more X-ray luminous 
than would be expected from LMXB emission alone, although several galaxies 
are marginally consistent with the LMXB relation. 

\begin{figure}
\figurenum{10} 
\plotone{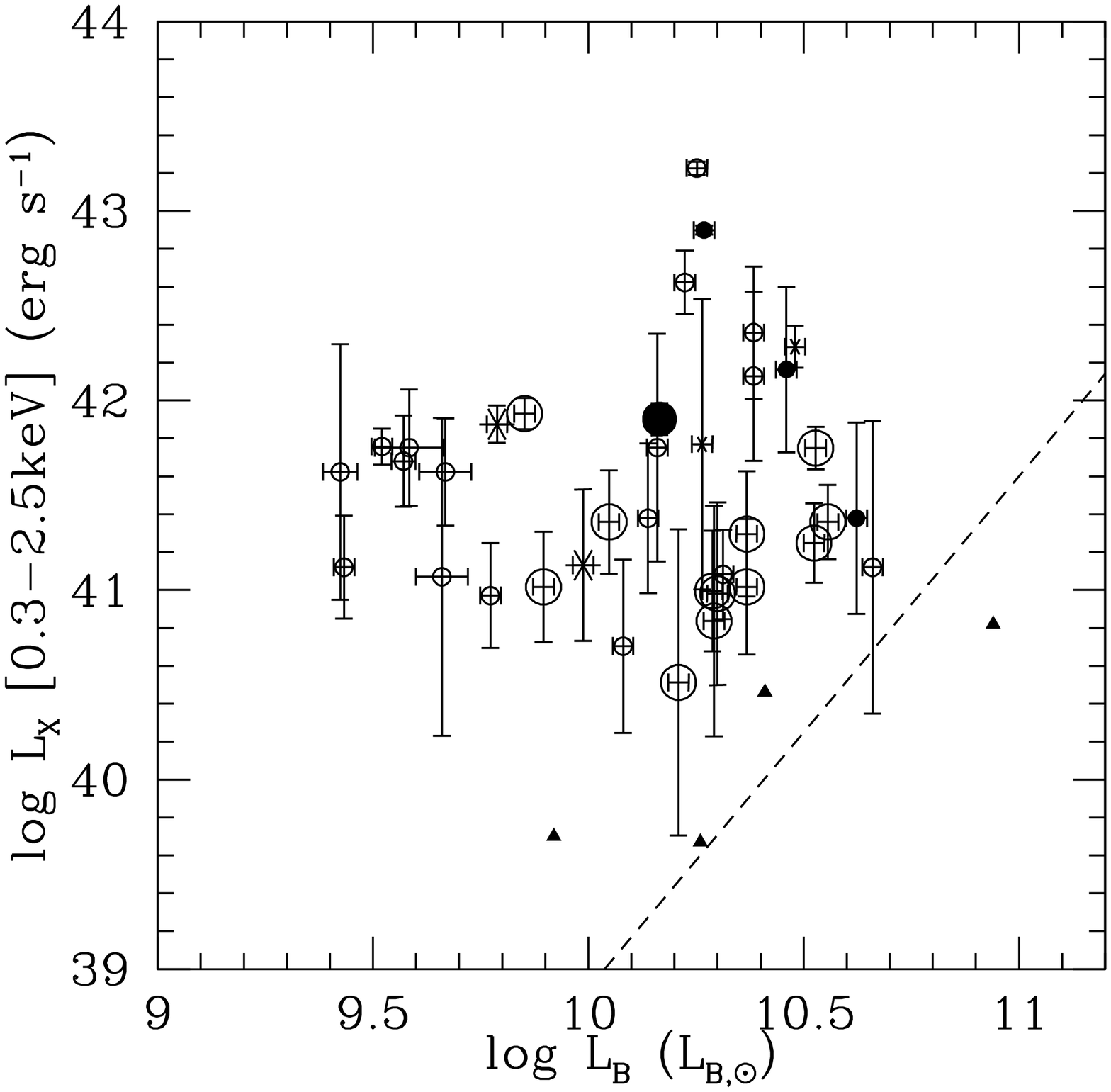} 
\caption{
Soft $L_X$ [0.3--2.5 keV] vs.\ $L_B$ for all of our sources with $B-$band 
photometry. The points are as in previous figures, while the solid line 
shows the relationship between B-band and soft X-ray luminosity found by 
\citet{osullivan03} for early-type galaxies with X-ray emission from hot 
gas. The filled triangles correspond to measurements of early-type 
galaxies in Abell~1367 by \citet{sun05}. Similar to Figure~\ref{fig:lxlb}, 
the soft X-ray luminosities of these sources are brighter for their B-band 
luminosity than expected if the emission was due to X-ray gas. 
\label{fig:lxlb2} 
}
\end{figure}

X-ray emission from the hot interstellar medium in early-type galaxies may 
also make a significant contribution to the total X-ray luminosity.  
The X-ray spectra of hot gas emission should be substantially softer than 
AGN emission, however most of the X-ray sources lack sufficient counts to 
use this as a discriminant. Instead, we compare these sources to the 
$L_X - L_B$ relation observed for local early-type galaxies 
\citep{matsushita01,yamasaki02}. Figure~\ref{fig:lxlb2} shows 
soft X-ray luminosity vs.\ $L_{B,\odot}$ for our sample along with 
the literature relation from \citet{osullivan03} and several measurements 
of early-type galaxies in Abell~1367 with \chandra\ from \citet{sun05}. 
The measurements of early-type in Abell~1367 galaxies are a particularly 
relevant comparison sample because the $L_X/L_B$ ratio is generally larger for 
galaxies in rich environments, provided the environments are not so rich 
and the galaxies have such high relative velocities that 
ram pressure stripping is significant \citep{brown00}. 

\section{Discussion} \label{sec:dis} 

All of our sources have at least seven counts in the broad (0.3-8 keV) 
X-ray band. Seven counts correspond to broad band X-ray luminosities 
ranging from $L_X = 2 \times 10^{40}$ \ergs (A644) to $L_X = 5 \times 10^{41}$ 
\ergs (A1689) for an unobscured $\Gamma = 1.7$ power law and Galactic 
absorption. When we combine the X-ray sensitivity with our spectroscopic 
completeness, we can conservatively state that we are sensitive to all X-ray 
sources more luminous than $L_X = 10^{42}$ \ergs and host galaxies with 
$M_R < -20$ mag in these eight clusters, although we are sensitive to 
substantially less X-ray luminous galaxies in most of the clusters. 

\subsection{Nature of the X-ray Emission} 

Only four X-ray sources in our sample (A644-1, A2104-1, A1689-2, A2163-1) are 
both X-ray luminous $L_X > 10^{42}$ \ergs\ and exhibit the emission-line 
signatures of AGN. Ten additional galaxies are above $L_X = 10^{42}$ \ergs\ 
and therefore are most likely AGN, but at progressively fainter X-ray 
luminosities an increasing number of physical mechanisms can produce luminous 
X-ray emission, particularly a population of LMXBs, thermal emission from a 
hot, gaseous halo, and star formation. Of these three mechanisms, we rule out 
star formation for these cluster sources because the observed \ha\ or \oii\ 
luminosities indicate low star formation rates that could not produce the 
observed X-ray luminosities \citep[e.g.][]{ranalli03,hornschemeier05}.  

While the less luminous sources lack sufficient counts for detailed 
spectral fitting, most with at least moderate signal-to-noise ratio 
X-ray detections are consistent with a $\Gamma = 1.7$ power law 
(see Figures~\ref{fig:xccd} and \ref{fig:xq}). The sources with 
softer X-ray spectra could be due to thermal bremsstrahlung or may be 
Compton-thick AGN. Our strongest arguments against emission from 
LMXBs or hot, gaseous halos are the ratios of the X-ray to B-band 
luminosities of the galaxies. 
From Figures~\ref{fig:lxlb} and \ref{fig:lxlb2}, all but three sources are 
more X-ray luminous than either the $L_X - L_B$ relation for LMXBs or hot 
halos (two of the three are consistent with both relations), or 34 of 37 
sources with visible-wavelength photometry.  
As noted above in section~\ref{sec:xdata}, all of these sources are 
unresolved, which strengthens the argument against an LMXB or hot halo origin 
for the lower-redshift clusters. 

There are three additional cluster X-ray sources that are not in the figures 
because we do not have B-band observations. Two of the three were detected 
in the hard X-ray band and one with sufficient counts to determine that it 
is consistent with an AGN. If we conservatively assume the other two 
sources are not AGN, along with the three mentioned previously that are 
statistically consistent with the LMXB or hot gas $L_X - L_B$ relation, we 
conclude that at least 35 of these 40 X-ray sources are AGN. 

\subsection{AGN Fraction in Clusters} 

We calculate the fraction $f_A$ of cluster galaxies that host AGN by dividing 
the number of cluster AGN $N_A$ by an estimate of the total number of cluster 
members $N_C$ to our absolute-magnitude spectroscopic limit $M_R$ 
within the field of the \chandra\ observation. 
Along with our spectroscopic survey of X-ray sources, we also obtained 
spectra of many candidate cluster members to measure the cluster velocity 
dispersion and more generally characterize the cluster galaxy population. 
We first calculate the fraction $f_C$ of all (non X-ray) spectroscopic 
targets to some magnitude limit that are cluster members. Then the 
total number of cluster galaxies to this magnitude limit is 
$N_C = f_C \times N_T$, the magnitude limit corresponds to the same 
$M_R$ for all clusters, and $N_T$ is the total galaxy population to this 
magnitude limit.  The cluster AGN fraction is then $f_A(<M_R) = N_A/N_C$. 

We estimated the total number of galaxies in each cluster to a uniform 
absolute magnitude of $M_R = -20$ mag, which was set by the spectroscopic 
completeness of our observations of the more distant clusters. 
Our success rate at identifying cluster members varied from 19 to 44\% based 
on spectroscopic redshifts for between 12 and 54 cluster members not targeted 
as X-ray sources. These calculations yield an X-ray source fraction of 
$f_A(M_R<-20) \sim 6$\% for the seven new clusters and confirms our previous 
result based on just one cluster \citep{martini02}. 
The six cluster members fainter than $M_R = -20$ mag in Table~\ref{tbl:lum} 
are not included in this calculation. 

The AGN fraction is more meaningfully compared to other surveys with a 
constraint on AGN luminosity in addition to galaxy stellar luminosity. 
We use the X-ray sensitivity of our observations to constrain the AGN fraction 
to be the fraction of AGN above a broad-band X-ray luminosity of 
$L_X = 10^{41}$ \ergs. 
Nearly all of these X-ray observations would detect an X-ray source more 
luminous than this and all but four of our 40 X-ray sources are above this 
luminosity. The four sources eliminated by this luminosity cut include all 
but one of those least likely to be AGN based on X-ray to visible-wavelength 
flux ratio. We find that the fraction of galaxies brighter than $M_R = -20$ mag 
that host AGN more luminous than $L_X = 10^{41}$ \ergs\ is 
$f_A(M_R<-20$ mag$;L_X>10^{41}$\ergs$) \sim 5 \pm 1.5$\%. 
The quoted uncertainty corresponds to 90\% one-sided confidence limits 
calculated from Poisson statistics \citep{gehrels86}. 
We may have underestimated the true AGN fraction because our X-ray sensitivity 
is a factor of five above $L_X = 10^{41}$ \ergs\ for two clusters (A1689 
and MS1008) and therefore these clusters may contain additional AGN below 
our X-ray limit and more luminous than $10^{41}$ \ergs. The advantage 
of this lower luminosity limit is that it maximizes the number of 
X-ray sources in cluster members and the statistical significance of the 
detection. 

While there are only a small number of AGN per cluster, we do find evidence 
for substantial variation in the AGN fraction from cluster to cluster, most 
notably a substantially higher AGN fraction in Abell 3128 relative to 
Abell 644, even though the Abell 644 X-ray observations are more sensitive. 
We plan to explore this aspect of our data in a future paper 
(Martini \etal\ 2006, {\it in prep}) that will include a more detailed 
completeness study. 
We note uncertainties in the present completeness calculation may introduce 
systematic errors in the AGN fraction that are larger than the statistical 
errors. 

The AGN fraction we find in clusters is approximately a factor of five times 
higher than the visible-wavelength spectroscopic survey of \citet{dressler85}, 
who found a $\sim 1$\% AGN fraction \citep[see also][]{dressler99}. 
However, when we just calculate the cluster AGN fraction from those 
identified in visible-wavelength spectroscopy, our results are completely 
consistent with the \citet{dressler85} result. This has some interesting 
implications for the field AGN fraction, which these authors found was 
approximately five times higher than the cluster fraction. 
If the properties of cluster and field AGN are similar, then there are simply 
five times fewer AGN in clusters. A survey with similar selection criteria 
to this one should then identify a field AGN fraction of $\sim 25$\%. 

\subsection{Contribution to the X-ray Source Density} 

We can also use these data to address the origin of the apparent 
excess of X-ray sources observed toward the fields of other clusters 
observed by \chandra. The excesses take the form of overdensities in 
the log N -- log S relation and have been reported in the soft 
\citep{ruderman05} and hard X-ray bands \citep[e.g.][]{cappelluti05}, 
although not for all cluster fields \citep{kim04b}.  
Our redshift measurements allow us to directly address whether or not the X-ray 
sources in these eight clusters add appreciably to the X-ray surface density.  
We address this question on a cluster by cluster basis by measuring the 
flux-limited fraction of sources that are members of a given cluster. For 
each of our fields, we have counted the number of sources brighter than some 
limiting soft X-ray flux and calculated the fraction of these sources 
that are cluster members. This effectively integrates log N -- log S for the 
field to the flux limit of the observation, where we have adopted a flux limit 
corresponding to five counts in the soft X-ray band at the aimpoint and 
adopted a model with $\Gamma = 1.7$ and only Galactic absorption. 
The five count soft-band threshold was chosen because it is comparable in 
signal-to-noise ratio to our seven count broad-band threshold. This threshold 
includes all but one of our cluster sources (MS1008-1). 

We find that the cluster X-ray sources contribute up to 20\% of the 
soft X-ray point source density in these fields. For each field we 
use Poisson statistics \citep{gehrels86} to calculate the one-sided 
confidence limits from the number of nonmembers. For three of these 
eight clusters, the point sources due to cluster members constitute a 
$\sim 1\sigma$ excess, and are slightly more significant for a 
fourth (A3128). These marginally significant excesses are similar to 
the results toward cluster fields without membership information, 
particularly the \citet{ruderman05} study that identified an $8\sigma$ 
excess surface density from a sample of 51 clusters. 
Two differences between our data and their MACS sample are 
that the later data have a shallower luminosity threshold and encompass a 
larger fraction of the total cluster size because the MACS clusters tend to be 
at higher redshift. 
The absence of an overdensity of X-ray sources in the \citet{kim04b} study, 
in contrast, may be due to their preference for clusters whose intracluster 
medium subtends only a small fraction (less than 10\%) of the field 
\citep{kim04a} with the result that their sample of 29 cluster fields is 
dominated by high-redshift clusters. Only the most luminous cluster members 
could therefore be detected in these images. In addition, as \citet{kim04b} 
note, their cluster fields were mostly obtained with ACIS-I and any cluster 
signal may also be diluted by field contamination due to the large field 
of view. 

\begin{figure*}
\figurenum{11} 
\plotone{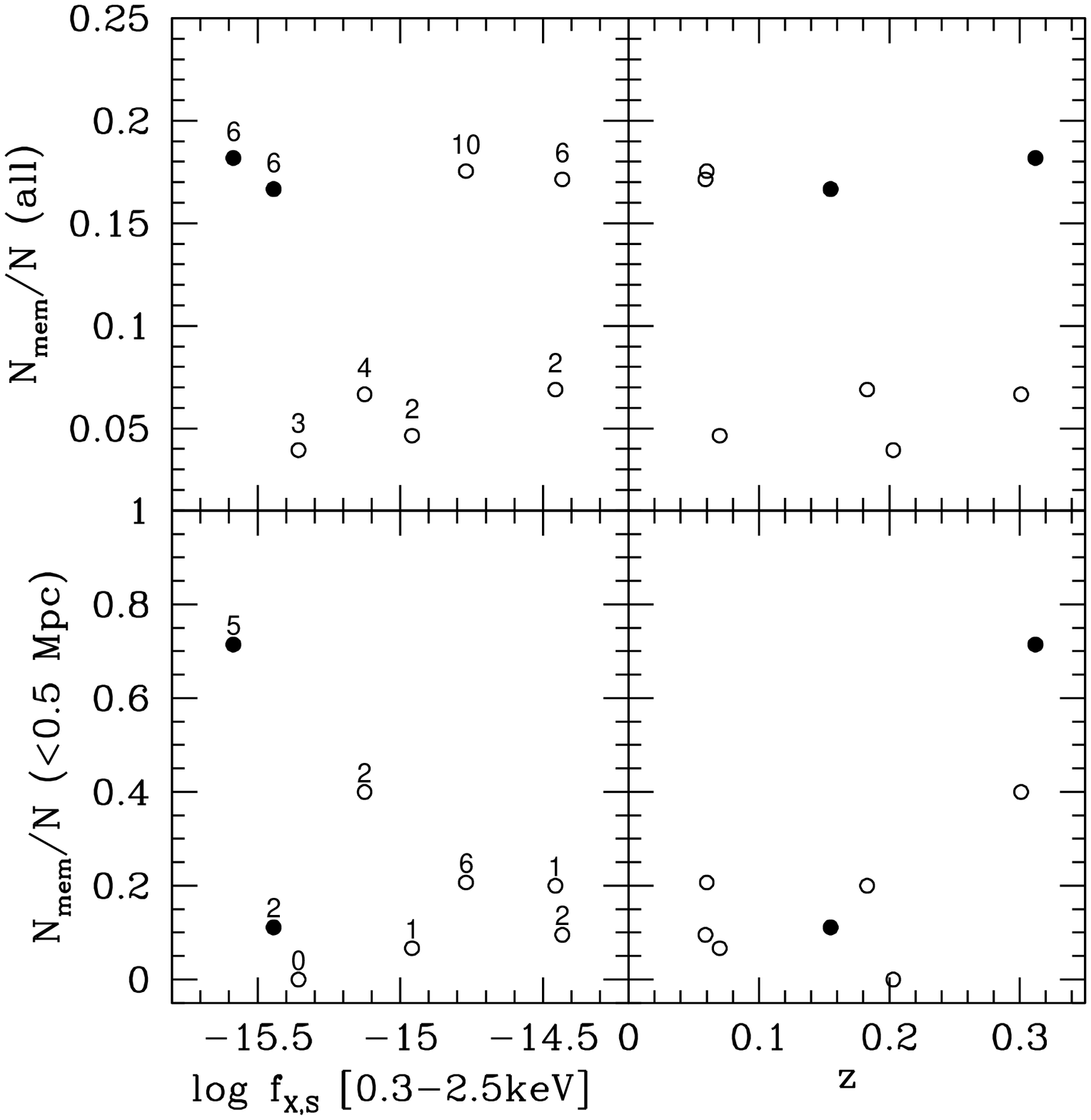} 
\caption{
Fraction of soft X-ray sources in each field due to cluster members as 
a function of limiting X-ray flux (5 soft counts, upper left panel) and 
redshift (upper right panel). The information is also shown for only 
sources within 0.5~Mpc of the \chandra\ aimpoint in the lower two 
panels. 
Clusters observed with the front-illuminated ACIS-I chips are plotted as open 
circles, while clusters observed with the back-illuminated ACIS-S chip are 
plotted as filled circles. 
The numbers above the points in the two left panels are the number of 
cluster members. 
\label{fig:logns} 
}
\end{figure*}

The redshift range and various sensitivity limits of these eight clusters
can be used to explore the importance of these quantities in detecting 
X-ray source excesses in other cluster fields. 
In the top panels of Figure~\ref{fig:logns} we show that there are no obvious 
trends with either flux limit or redshift, in spite of the approximately 
factor of ten variation in sensitivity. This lack of variation may be due in 
part to the fact that the more distant cluster observations were chosen to 
be more sensitive and therefore the observations have a narrower range in 
luminosity sensitivity than flux sensitivity, although this could be offset 
by higher field contamination at fainter flux limits. Indeed three of 
the four clusters with marginally-significant excesses, which correspond 
to those with greater than 10\% cluster contributions, are at $z < 0.1$. 
We note that we are neglecting the fact that the flux limit will rise with 
increasing off-axis angle because both the cluster and field X-ray sources 
will experience the same effect, although could nevertheless be a 
bias toward a higher cluster fraction if the cluster sources are 
more centrally concentrated within the ACIS field of view. 

The radial distribution of cluster sources is also important because the fixed 
angular size of the \chandra\ observations subtends a substantially smaller 
fraction of a given cluster at lower-redshift. 
For example, the approximately $17'$ wide ACIS-I field of view corresponds to 
only 1.2~Mpc at $z=0.06$, but over 4.5~Mpc at $z=0.3$. 
We have investigated the importance of aperture size by calculating the 
fraction of cluster X-ray sources within a fixed physical radius of 0.5~Mpc 
for all eight clusters. 
This physical aperture size is the largest that is contained within the 
\chandra\ images of all eight clusters, particularly the two lowest-redshift 
clusters, and also happens to be the physical radius within which 
\citet{ruderman05} observed the most significant overdensity of X-ray sources. 
Within this aperture size we now measure fractions as high as $\sim 70$\% for 
one cluster (AC114), although we still measure $\sim 20$\% or less for most 
(six clusters; all but MS1008 and AC114)
The fractional contribution of sources within 0.5~Mpc as a function of limiting 
soft X-ray flux and redshift are shown in the bottom panels of 
Figure~\ref{fig:logns}. 
If we now repeat the same experiment described above to determine if 
the cluster members make a statistically-significant contribution to the 
surface density within an 0.5~Mpc aperture, the cluster X-ray sources in 
A3128 and AC114 are $>1\sigma$ excesses over the number of other X-ray 
sources, while MS1008 sources make only a marginal contribution. 
We note that the $8\sigma$ excess noted by \citep{ruderman05} in their 
sample of 51 clusters is consistent with a $\sim 1\sigma$ excess per 
cluster summed in over their sample. 
While the fraction of cluster members does appear highest for the two 
highest-redshift clusters, the present sample of clusters is too small 
for us to claim the presence of redshift dependence in the cluster 
AGN population. 

\section{Summary} \label{sec:sum} 

We have completed a redshift survey and multiwavelength study of X-ray sources 
in the fields of eight clusters of galaxies and discovered luminous 
X-ray emission from 40 galaxies. Emission-lines, X-ray spectral properties, 
and X-ray to visible-wavelength flux ratios were used to determine that 
the vast majority of these sources are low-luminosity AGN. We conservatively 
estimate that at least 35 of these 40 sources are AGN in the clusters and 
estimate that the AGN fraction in clusters is 
$f_A(M_R<-20;L_X>10^{41}) \sim 5$\%, that is $\sim 5$\% of all cluster galaxies 
more luminous than $M_R = -20$ mag host AGN with broad-band X-ray luminosities 
greater than $L_X = 10^{41}$ \ergs. 

Only a small fraction of these galaxies would be classified as AGN from their 
visible-wavelength spectra. In most instances the AGN identification 
is due to X-ray spectral shape or X-ray to visible-wavelength flux ratios. 
In particular, the multiwavelength properties of these sources are used 
to show that they are more X-ray luminous for their observed B-band 
luminosity than would be expected for either a population of LMXBs or 
hot, gaseous halos that have survived in the intracluster medium. 

The X-ray sources in these clusters do not make a statistically significant 
contribution to the surface density in any one of these \chandra\ 
observations, although summation of a sufficient number of similar clusters 
would produce a statistically significant excess similar to those 
observed in stacked \chandra\ observations of other cluster fields. 
We therefore conclude there is a sufficiently large population of X-ray 
sources in clusters of galaxies to explain the observed point source excess.  

This population of cluster AGN is a factor of five higher than expected 
from previous spectroscopic surveys. We attribute our higher success rate 
to the fact that optical searches for AGN are strongly biased against 
detection of low-luminosity AGN because of host galaxy dilution and 
obscuration. 
AGN with yet lower X-ray luminosities are likely present in other luminous 
cluster galaxies, or equivalently the AGN fraction should increase with a 
lower $L_X$ threshold, although they may prove difficult to discriminate from 
emission due to LMXBs or hot gas in the absence of substantially higher 
signal-to-noise ratio data. 
Our result also has interesting implications for a higher field AGN fraction. 
If the AGN fraction in the field is five times the cluster AGN fraction, as 
found by Dressler and collaborators for uniformly-selected samples of field 
and cluster galaxies, then similar selection criteria to this study will 
identify a field AGN fraction of $\sim 25$\%. 

The substantial number of cluster galaxies hosting AGN suggests that cluster 
galaxies can retain significant reservoirs of cold gas near their central, 
supermassive black holes. These AGN, and others currently dormant, may have 
significantly contributed to heating the intracluster medium and driving 
galaxy evolution via feedback processes at higher redshift. Future 
observations of AGN and measurements of the AGN fraction in higher-redshift 
clusters may provide valuable new insights into cluster assembly and galaxy 
evolution in clusters. 

\acknowledgements 

PM was supported by a Carnegie Starr Fellowship and a CfA Clay Fellowship. 
Support for this work was provided by the National Aeronautics and Space 
Administration through Chandra Award Numbers 04700793 and 05700786 issued by 
the Chandra X-ray Observatory Center, which is operated by the Smithsonian 
Astrophysical Observatory for and on behalf of the National Aeronautics 
Space Administration under contract NAS8-03060.
We greatly appreciate the excellent staffs of the Las Campanas Observatory 
and the Magellan Project Telescopes for their assistance with these 
observations.
We thank Tom Aldcroft for help with X-ray analysis software during an earlier 
stage of this project and Mark Whittle and the referee for helpful 
suggestions and comments. 
The Canadian Astronomy Data Centre, which is operated by the Dominion 
Astrophysical Observatory for the National Research Council of Canada's 
Herzberg Institute of Astrophysics, was an extremely helpful source for 
our photometric calibration. 
This research has made use of the NASA/IPAC Extragalactic Database (NED) 
which is operated by the Jet Propulsion Laboratory, California Institute of 
Technology, under contract with the National Aeronautics and Space 
Administration.
The Digitized Sky Surveys were produced at the Space Telescope Science 
Institute under U.S. Government grant NAG W-2166. The images of these surveys 
are based on photographic data obtained using the Oschin Schmidt Telescope on 
Palomar Mountain and the UK Schmidt Telescope. The plates were processed into 
the present compressed digital form with the permission of these institutions.

\end{document}